%% file: main.tex
\documentclass[letterpaper,twocolumn,10pt]{article}
\usepackage{usenix,epsfig}
\usepackage{xcolor}
\usepackage{colortbl}
\usepackage{booktabs}
\usepackage{makecell}
\usepackage{multirow}
\usepackage{pifont}
\usepackage{tikz}
\usetikzlibrary{shapes.geometric, calc}
\usepackage{amsmath}
\usepackage{bbm}
\usepackage{cite}
\usepackage{graphicx}
\usepackage{array}
\usepackage{enumitem}        
\usepackage{float}           
\usepackage{xspace}          
\input{styles.tex}           

\begin{document}

\date{}


\title{\Large \bf GlitchLab: A Hardware-in-the-Loop Optimizer for Physical Fault Injection}


\author{
{\rm Tanvir Hossain \qquad Abhinav Mahadevan \qquad Jasper Van Woudenberg}\\
{\rm Rajesh Velegalati \qquad Arindam Bhattacharyya}\\
Keysight Technologies, Inc.
}

\maketitle

\thispagestyle{empty}

\subsection*{Abstract}

Physical fault injection can turn brief hardware disturbances into security failures such as key recovery, authentication bypass, and unintended control flow. Finding effective faults is difficult because many interacting parameters create a large search space, successful settings are sparse and target-dependent, and each hardware attempt provides limited feedback. Under fixed testing time, efficient search is therefore critical for assessing fault sensitivity.

We present \textsc{GlitchLab}, an online hardware-in-the-loop
platform that treats delay as a timing gate, voltage and pulse duration as
severity controls, and hardware outcomes as structured feedback. It implements
RL-Q (Q-learning-based reinforcement learning), a structured bandit for
discovery, and Structured-Outcome-Based Adaptive Search (SOBAS), a model-based
policy for fault reproduction.

Both policies find a target fault in every AES, password, and
control-flow campaign. On AES and control flow, they require $2$--$85\times$
fewer attempts and $26$--$1{,}237\times$ less time than the baselines; on
password, both succeed while the baselines fail within $5{,}000$ attempts.
After discovery, SOBAS reproduces faults $7.3$--$21\times$ more often, while
RL-Q identifies $30\%$ more distinct AES settings.

\input{sections/introduction}
\input{sections/background}

\input{sections/threat_model}
\input{sections/method}
\input{sections/experimental_setup}

\input{sections/evaluation}
\input{sections/security_impact}
\input{sections/discussion}

\input{sections/conclusion}


{\footnotesize \bibliographystyle{unsrt}
\bibliography{references}}

\appendix
\input{sections/appendix}

\input{sections/open_science_appendix}
\end{document}

%% file: styles.tex
\definecolor{ksRed}{RGB}{233,0,41}          
\definecolor{ksDarkRed}{RGB}{180,20,40}     
\definecolor{ksCharcoal}{RGB}{84,86,90}     
\definecolor{ksGray}{RGB}{120,120,125}      
\definecolor{ksSlate}{RGB}{90,100,120}      
\definecolor{ksAmber}{RGB}{180,110,40}      
\definecolor{ksNeutral}{RGB}{140,140,145}   
\definecolor{ksLightGray}{RGB}{231,231,234} 
\definecolor{ksMedGray}{RGB}{165,165,165}   
\definecolor{ksBlack}{RGB}{30,30,32}        

\definecolor{ksPink}{RGB}{190,60,120}       
\definecolor{ksYellow}{RGB}{165,135,15}     
\definecolor{ksGreen}{RGB}{40,120,65}       
\definecolor{ksMagenta}{RGB}{140,40,120}    

\definecolor{tblhead}{RGB}{240,240,242}     
\definecolor{tblours}{RGB}{255,240,240}     

\definecolor{linkRed}{RGB}{180,20,40}
\definecolor{linkConcept}{RGB}{200,30,50}
\definecolor{linkStat}{RGB}{70,80,90}

\DeclareRobustCommand{\ptag}[2]{%
  \tikz[baseline=(X.base)]{%
    \node[
      draw=#1!70,
      fill=#1!12,
      line width=0.4pt,
      rounded corners=2.2pt,
      inner xsep=3.8pt,
      inner ysep=1.6pt,
      font=\sffamily\bfseries\scriptsize
    ] (X) {\textcolor{#1}{#2}};
  }%
}

\DeclareRobustCommand{\GlitchLab}{\ptag{ksRed}{GlitchLab}\xspace}
\DeclareRobustCommand{\RED}{\ptag{ksRed}{RED}\xspace}          

\DeclareRobustCommand{\RLQ}{\ptag{ksCharcoal}{RL-Q}\xspace}
\DeclareRobustCommand{\BOGP}{\ptag{ksCharcoal}{BO-GP}\xspace}
\DeclareRobustCommand{\BOSMAC}{\ptag{ksCharcoal}{BO-SMAC}\xspace}
\DeclareRobustCommand{\BOSOBAS}{\ptag{ksCharcoal}{SOBAS}\xspace}

\DeclareRobustCommand{\AES}{\ptag{ksGray}{AES}\xspace}
\DeclareRobustCommand{\FiniteLoop}{\ptag{ksGray}{finite-loop}\xspace}

\DeclareRobustCommand{\OutcomePink}{\ptag{ksPink}{Pink}\xspace}
\DeclareRobustCommand{\OutcomeYellow}{\ptag{ksYellow}{Yellow}\xspace}
\DeclareRobustCommand{\OutcomeGreen}{\ptag{ksGreen}{Green}\xspace}
\DeclareRobustCommand{\OutcomeMagenta}{\ptag{ksMagenta}{Magenta}\xspace}

\newcommand{\yescheck}{\textcolor{ksCharcoal}{\scalebox{1.12}{\ding{51}}}}
\newcommand{\warntri}{\textcolor{ksAmber}{\scalebox{1.0}{\ding{115}}}}
\newcommand{\nocross}{\textcolor{ksRed}{\scalebox{1.12}{\ding{55}}}}
\newcommand{\nostated}{\textcolor{ksNeutral}{\scalebox{1.12}{--}}}

\DeclareRobustCommand{\ResultIconDiscovery}{%
  \tikz[baseline=-0.55ex,x=1pt,y=1pt]{%
    \draw[ksDarkRed,line width=.55pt] (0,0) circle (4.7);
    \draw[ksDarkRed,line width=.55pt] (-6.8,0)--(6.8,0) (0,-6.8)--(0,6.8);
    \fill[ksRed] (0,0) circle (1.35);}}
\DeclareRobustCommand{\ResultIconWindow}{%
  \tikz[baseline=-0.55ex,x=1pt,y=1pt]{%
    \draw[ksDarkRed,line width=.55pt] (-7,0)--(7,0);
    \draw[ksDarkRed,line width=.55pt] (-7,-4)--(-7,4) (7,-4)--(7,4);
    \draw[ksRed,line width=2.1pt] (-2.2,0)--(2.2,0);}}
\DeclareRobustCommand{\ResultIconGeometry}{%
  \tikz[baseline=-0.55ex,x=1pt,y=1pt]{%
    \draw[ksDarkRed,line width=.55pt] (-6,-5)--(-6,5) (-6,-5)--(7,-5);
    \draw[ksRed,line width=.9pt] (-5,-4) .. controls (-1,-3) and (1,2) .. (6,4);}}
\DeclareRobustCommand{\ResultIconYield}{%
  \tikz[baseline=-0.55ex,x=1pt,y=1pt]{%
    \draw[ksDarkRed,line width=.55pt] (-6,-5)--(-6,5) (-6,-5)--(7,-5);
    \fill[ksRed!45] (-4,-5) rectangle (-1,-1);
    \fill[ksRed!70] (0,-5) rectangle (3,2);
    \fill[ksRed] (4,-5) rectangle (7,5);}}
\DeclareRobustCommand{\ResultIconCoverage}{%
  \tikz[baseline=-0.55ex,x=1pt,y=1pt]{%
    \draw[ksDarkRed,line width=.5pt] (-6,-6) grid[step=4pt] (6,6);
    \fill[ksRed!45] (-6,2) rectangle (-2,6);
    \fill[ksRed!70] (-2,-2) rectangle (2,2);
    \fill[ksRed] (2,-6) rectangle (6,-2);}}

\DeclareRobustCommand{\ResultBox}[2]{%
  \par\begingroup
  \setlength{\fboxsep}{2pt}%
  \setlength{\fboxrule}{.45pt}%
  \noindent\fcolorbox{ksRed!60!ksBlack}{ksLightGray}{%
    \parbox{\dimexpr\columnwidth-2\fboxsep-2\fboxrule\relax}{%
      \small\setlength{\parindent}{0pt}\setlength{\parskip}{0pt}%
      \hangindent=1.72em\hangafter=1%
      \makebox[1.45em][c]{#1}\hspace{.27em}\color{ksBlack}#2%
    }}%
  \endgroup\par\vspace{-2pt}}

\providecommand{\xspace}{\space}

\floatstyle{ruled}
\newfloat{algorithm}{tbp}{loa}
\floatname{algorithm}{Algorithm}
\floatstyle{plain}                       

\newlength{\algindent}
\newcounter{alglineno}
\newcommand{\algbody}{%
  \par\vspace{0.4ex}\setcounter{alglineno}{0}%
  \begingroup\footnotesize\raggedright%
  \setlength{\parindent}{0pt}\setlength{\parskip}{0.15ex}%
}
\newcommand{\algend}{\endgroup\par\vspace{0.3ex}}
\newcommand{\aline}[2]{%
  \refstepcounter{alglineno}%
  \noindent\makebox[1.5em][r]{\scriptsize\textcolor{ksMedGray}{\thealglineno:}}%
  \hspace{0.45em}\hspace{#1\algindent}%
  \hangindent=\dimexpr 1.95em + #1\algindent\relax
  \ignorespaces #2\par}
\newcommand{\kw}[1]{\textbf{#1}}
\newcommand{\cmt}[1]{{\itshape\textcolor{ksCharcoal}{$\triangleright$ #1}}}

\newsavebox{\cwrefbox}

\providecommand{\todo}[1]{}


%% file: sections/introduction.tex
\section{Introduction}\label{sec:introduction}

Physical fault injection (FI) is a practical threat to embedded systems because
a precisely timed perturbation can corrupt a security-critical computation
without exploiting a software vulnerability~\cite{fi_survey,yuce2018faults}.
Voltage and clock tampering, electromagnetic pulses, and laser pulses
have all induced computational faults in security-relevant
devices~\cite{korak2014tampering,dehbaoui2012emfi,moro2013emfi,
skorobogatov2003optical}. Demonstrated consequences include cryptographic key
recovery, control-flow redirection, privilege escalation, and integrity
violations~\cite{boneh1997faults,biham1997dfa,piret2003dfa,timmers2016pc,
timmers2017linux,tang2017clkscrew,chen2021voltpillager}. Executing an injection is straightforward once the
bench is configured; finding an effective configuration is not. The
perturbation must overlap the vulnerable operation and be strong enough to
change architectural state without only resetting the device or disrupting
communication. Consequently, fault injection is also an online experimental
search problem.

A voltage-FI campaign selects a glitch voltage $v$, trigger-relative delay $d$,
and pulse duration $\ell$. Exhaustive and random search allocate attempts
without using the outcomes already observed~\cite{bergstra2012random}.
Sequential model-based and Bayesian optimization instead use observations to
select subsequent evaluations~\cite{jones1998ego,snoek2012bo,frazier2018bo,
bayesian_optimization}, but two sources of information remain underused in FI.
First, a hardware
attempt returns more than a binary success bit: normal execution, a malformed
response, a non-target fault, reset, synchronization failure, and a target
fault imply different properties of the tested configuration. Second, the
search dimensions have distinct roles: coarse delay selects the
operation, fine delay selects the clock phase, and voltage and pulse duration
set perturbation severity (Section~\ref{sec:rlq-timing}). Treating these
dimensions or their outcomes as interchangeable discards useful feedback.

This observation leads to a campaign-level problem formulation. Given bounds
$\mathcal{X}$, no prior fault map, and a budget of $B$ hardware attempts, the
policy must first minimize the attempt index of the first target fault. After
discovery, the appropriate objective depends on the security question: an
attacker may seek a reproducible target effect, whereas an evaluator may need
to characterize the extent of the vulnerable region. These goals are related
but not equivalent. A policy that repeatedly samples one high-probability
configuration can have high target-fault yield and poor region coverage; a
policy that traces the fault boundary can have the opposite behavior.

Our central contribution is an online, hardware-in-the-loop search
method that reaches target faults using far fewer hardware attempts than
existing baselines. It treats delay, voltage, and pulse length as physically
distinct search roles and uses the full structured hardware response instead
of a binary success bit, a structure that extends naturally to other FI
modalities (e.g. EMFI). We implement it with two complementary policies. \RLQ{} uses a
reset-based structured bandit derived from tabular Q-value
estimation~\cite{watkins1992qlearning,sutton2018rl}. It
allocates attempts across voltage--delay regions, brackets the pulse-length
transition, and uses each outcome class to update the corresponding physical
estimate. \BOSOBAS{} (Structured-Outcome-Based Adaptive Search) makes fewer
geometric assumptions: random-forest classifiers~\cite{breiman2001rf} estimate
categorical outcome probabilities, and the acquisition policy switches from boundary exploration
to target-fault acquisition after discovery. We implement both
policies in the commercial \GlitchLab{} platform, which provides a closed-loop
path for candidate selection, instrument execution, target recovery, outcome
classification, and logging.

Evaluation covers three physical units of the same Riscure Pinata model and three
classes of program: cryptographic computation, authentication, and control
flow. \RLQ{} and \BOSOBAS{} find a target fault in every campaign, using
$2$--$85\times$ fewer hardware attempts than the faster reference
configuration, while both \BOGP{} (Bayesian optimization with a
Gaussian-process surrogate) and \BOSMAC{} (SMAC, a random-forest-based
configurator) exhaust the shared budget on
the authentication target. After discovery the two policies separate:
\BOSOBAS{} reproduces the target outcome $7.3$--$21\times$ more often, and
\RLQ{} covers $30\%$ more of the AES voltage--length plane at equal yield.
The two policies therefore change places depending on which of these
quantities a campaign chooses to report, on the same hardware and the same
budget. A report built on one metric can consequently recommend the wrong
policy. Section~\ref{sec:security} then shows that the faults located here
support three attacks: AES key recovery, control-flow subversion, and
authentication bypass.

\paragraph{Contributions.}

This paper makes the following contributions:

\begin{itemize}[leftmargin=1.2em,itemsep=2pt,topsep=2pt,parsep=0pt,partopsep=0pt]
    \item \textbf{An online search that exploits distinct parameter
    roles and structured outcomes.} We treat delay as a timing gate and
    voltage/pulse length as severity controls rather than three equivalent
    black-box dimensions, and use the full five-class hardware response instead
    of a binary success bit, reaching target faults with $2$--$85\times$ fewer
    attempts than established baselines. Within this formulation, we separate
    first-target discovery, fault reproduction, and region characterization as
    distinct campaign objectives, enabling objective-specific policy design and
    comparison under one protocol.

    \item \textbf{Two search policies that use the full hardware response.}
    \RLQ{} exploits FI structure through delay localization, region allocation,
    and pulse-length bracketing. \BOSOBAS{} instead learns categorical outcome
    models and requires no explicit fault-boundary geometry. Both policies run
    end to end in \GlitchLab{} through common instrument, oracle, recovery, and
    logging interfaces. \textit{Although evaluated on voltage fault injection, the
    formulation can extend to other bounded search spaces when the platform
    exposes modality-specific controls and informative outcome feedback}.

    \item \textbf{Controlled hardware evidence of objective-dependent
    rankings.}
    Both proposed policies find a target fault in every measured
    run on the three physical boards.
    Under matched conditions,
    \BOSOBAS{} reproduces target faults $7.3$--$21\times$ more often, whereas
    \RLQ{} covers $30\%$ more AES fault settings at equal yield. Thus, no single
    policy dominates discovery, reproduction, and characterization.

    \item \textbf{Validation on three security-relevant fault classes.}
    The located faults produce AES ciphertext differences suitable for
    differential fault analysis, incorrect-password acceptance, and premature
    loop termination. These outcomes connect the search objectives to concrete
    cryptographic, authentication, and control-flow security failures.
\end{itemize}


\paragraph{Ethics and Responsible Use.}
\vspace{-3pt}
All experiments use laboratory targets owned or controlled by the researchers.
The work reports no previously unknown vendor-specific vulnerability and
therefore did not trigger coordinated disclosure. We treat
\textsc{GlitchLab} as dual-use infrastructure intended for authorized
security evaluation.

\paragraph{Roadmap.}
\vspace{-3pt}
Sections~\ref{sec:background} and~\ref{sec:threat} establish the FI search
problem and threat model. Section~\ref{sec:design} presents the two policies and
their \GlitchLab{} integration. Section~\ref{sec:impl} describes the bench and
experimental protocol; Section~\ref{sec:eval} evaluates the policies,
Section~\ref{sec:security} relates the faults to security outcomes, and
Section~\ref{sec:discussion} interprets the resulting tradeoffs.

%% file: sections/background.tex
\section{Background and Related Work}\label{sec:background}

This section summarizes FI physics, campaign feedback, and prior
adaptive search methods used in Section~\ref{sec:design}.

\subsection{Physical Fault Injection}\label{sec:bg-fi}

\begin{figure}[t]

    \centering
    \includegraphics[width=\columnwidth]{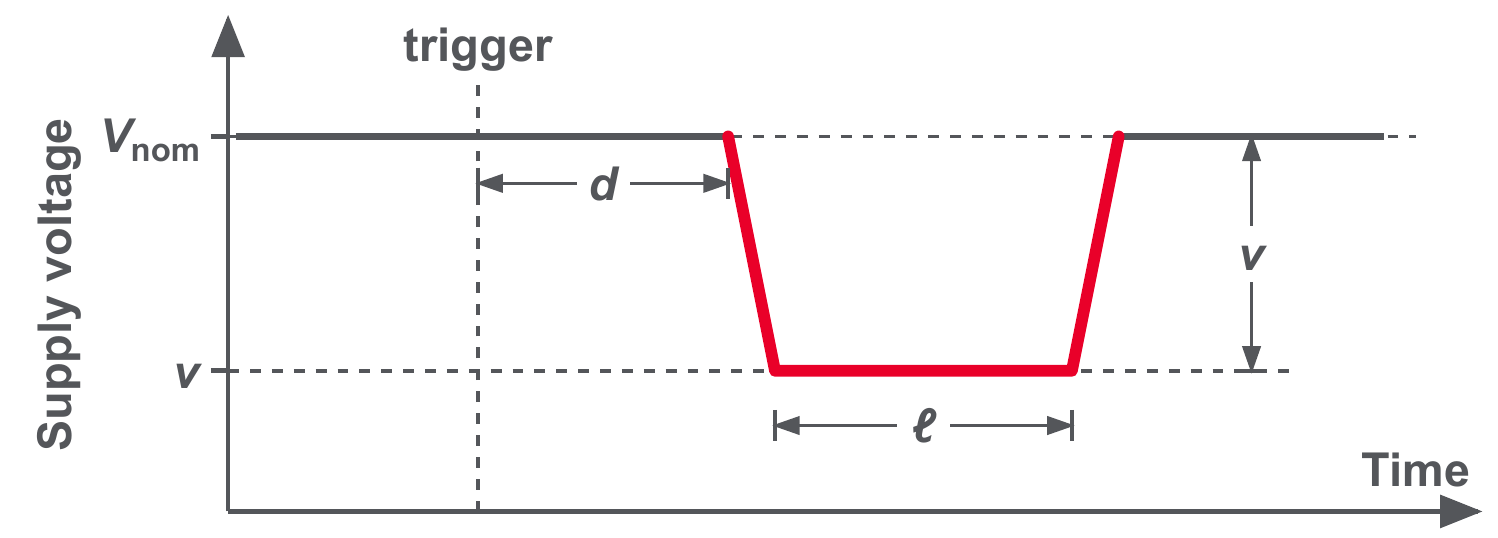}
    \vspace{-7pt}
    \caption{Voltage fault-injection parameters. The trigger-relative delay
    $d$ determines when the perturbation begins, the pulse duration $\ell$
    determines how long it persists, and $v$ denotes the glitch voltage.}
    \label{fig:voltage-glitch}
    \vspace{-12pt}
\end{figure}

Physical fault injection (FI) deliberately perturbs a device under test (DUT)
during execution to induce a computational error~\cite{fi_survey,
boneh1997faults,biham1997dfa,yuce2018faults}. Voltage glitching disturbs a
supply rail, clock glitching modifies the clock waveform, electromagnetic FI
couples a transient field into the circuit, and laser FI deposits energy at a
selected location. Each modality seeks a transient error that changes
architectural state without permanently damaging the
DUT~\cite{korak2014tampering,dehbaoui2012emfi,moro2013emfi,
skorobogatov2003optical}.

This work focuses on voltage fault injection (VFI). A glitching instrument
drives a supply rail away from its nominal voltage $V_{\mathrm{nom}}$ toward a
glitch voltage $v$, beginning at a trigger-relative delay $d$ and persisting
for a pulse duration $\ell$~\cite{korak2014tampering,
timmers2017linux}. We represent one glitch configuration as

\vspace{-7pt}
\begin{equation}
    \mathbf{x} = (v,d,\ell),
    \label{eq:glitch-configuration}
    \vspace{-9pt}
\end{equation}
\vspace{-7pt}

\noindent where $v$ determines the perturbation strength, $d$ aligns it with the
operation of interest, and $\ell$ determines its duration
(Figure~\ref{fig:voltage-glitch}). This is the VFI search vector; other modalities add controls such as EMFI probe position (Section~\ref{sec:discussion}). $\mathcal{W}\subseteq[d_{\min},d_{\max}]$ is the sensitive delay
interval, spanning the operation window and its clock-phase window.

The three parameters act jointly through the DUT's electrical and timing
behavior. Lower supply voltage reduces transistor drive strength, which can
increase propagation delay along active paths.
A sufficiently strong transient increases path delay beyond the
available timing margin, so an incorrect or unstable state is captured at a
clock edge~\cite{fi_survey,yuce2018faults}. The delay $d$ determines which
operation is active, while $v$ and $\ell$ determine whether the disturbance is
too weak to alter execution, strong enough to induce a useful fault, or severe
enough to crash or reset the device. The exact vulnerable region is difficult
to predict for commercial off-the-shelf devices whose physical implementation
and timing information are unavailable. Finding an effective combination of
voltage, timing, and pulse duration is therefore a search problem.

\subsection{Fault-Injection Campaigns and Parameter Search}
\label{sec:bg-campaigns}\label{sec:bg-oracles}

A practical FI campaign evaluates a sequence of configurations
$\mathbf{x}_1,\mathbf{x}_2,\ldots,\mathbf{x}_N$ within the bounded search space
$\mathcal{X}=[v_{\min},v_{\max}]\times[d_{\min},d_{\max}]
\times[\ell_{\min},\ell_{\max}]$. Each candidate requires one instrumented hardware attempt, including DUT
execution, response classification, and, when necessary, reset and recovery.
Policy computation is small relative to this experimental loop; consequently,
the number of hardware attempts is the primary campaign resource.

\begin{table*}[!t]
    \vspace{-4pt}
    \centering
    \caption{Comparison of adaptive parameter-search methods for physical
    fault injection.}\label{tab:related}
    {\scriptsize
    \setlength{\tabcolsep}{2.0pt}
    \renewcommand{\arraystretch}{1.22}
    \begin{tabular}{@{}>{\raggedright\arraybackslash}p{0.100\textwidth}
                       >{\raggedright\arraybackslash}p{0.046\textwidth}
                       >{\raggedright\arraybackslash}p{0.118\textwidth}
                       >{\raggedright\arraybackslash}p{0.092\textwidth}
                       !{\color{ksMedGray}\vrule width 0.7pt}
                       *{10}{c}
                       !{\color{ksMedGray}\vrule width 0.7pt}
                       >{\raggedright\arraybackslash}p{0.082\textwidth}@{}}
        \arrayrulecolor{ksBlack}\toprule
        \multicolumn{4}{@{}c!{\color{ksMedGray}\vrule width 0.7pt}}{%
            \cellcolor{tblhead}\footnotesize\bfseries\textcolor{ksCharcoal}{Work description}} &
        \multicolumn{10}{c!{\color{ksMedGray}\vrule width 0.7pt}}{%
            \cellcolor{tblhead}\footnotesize\bfseries\textcolor{ksCharcoal}{Capability}} &
        \multicolumn{1}{c@{}}{%
            \cellcolor{tblhead}\footnotesize\bfseries\textcolor{ksCharcoal}{Objective}} \\
        \addlinespace[1.2pt]
        \bfseries Work & \bfseries Mode & \bfseries Search method
        & \bfseries Parameters
        & \makecell{\bfseries No prior\\\bfseries map}
        & \makecell{\bfseries Closed-\\\bfseries loop}
        & \makecell{\bfseries Multi-\\\bfseries outcome}
        & \makecell{\bfseries Explore/\\\bfseries exploit}
        & \makecell{\bfseries Fault\\\bfseries region}
        & \makecell{\bfseries Target\\\bfseries rate}
        & \makecell{\bfseries Timing\\\bfseries window}
        & \makecell{\bfseries Cross-\\\bfseries device}
        & \makecell{\bfseries Cross-\\\bfseries program}
        & \makecell{\bfseries Cold\\\bfseries start}
        & \\
        \arrayrulecolor{ksRed}\specialrule{0.9pt}{0pt}{0pt}
        \rowcolor{tblours}
        \textcolor{ksRed}{\bfseries This work}
        & VCC & \textbf{RL-Q} and \textbf{SOBAS}
        & $v,d,\ell$
        & \yescheck & \yescheck & \yescheck & \yescheck & \yescheck
        & \yescheck & \yescheck
        & \scalebox{1.0}{\ding{115}}
        & \yescheck & \yescheck
        & First-target attempts; swappable policies \\
        \arrayrulecolor{ksRed}\specialrule{0.8pt}{0pt}{0pt}
        \arrayrulecolor{ksMedGray!45}
        Boix Carpi et al.\newline CARDIS'13~\cite{boixcarpi2013glitch}
        & VCC & Monte Carlo; FastBoxing; zoom-and-bound; genetic algorithm
        & Voltage, length; second-stage timing
        & \yescheck & \yescheck & \yescheck & \warntri & \yescheck
        & \warntri & \nocross & \yescheck & \warntri & \yescheck
        & Successful FI; bound region of interest \\
        \hline
        \rowcolor{ksLightGray!55}
        Maldini et al.\newline 2019~\cite{maldini2019optimizing}
        & EM & Genetic algorithm
        & 5 EMFI parameters
        & \yescheck & \yescheck & \warntri & \warntri & \nostated
        & \nostated & \nocross & \warntri & \yescheck & \yescheck
        & Successful EMFI; SHA-3 and PIN \\
        \hline
        Bozzato et al.\newline TCHES'19~\cite{bozzato2019shaping}
        & VCC & Arbitrary-waveform design; campaign-driven
        & Waveform shape; timing
        & \yescheck & \nocross & \nostated & \nostated & \nostated
        & \nostated & \nocross & \yescheck & \warntri & \yescheck
        & Firmware extraction \\
        \hline
        \rowcolor{ksLightGray!55}
        Werner et al.\newline CARDIS'21~\cite{werner2021fast}
        & VCC & SMAC (random-forest surrogate); successive halving;
          genetic algorithm
        & 9 waveform parameters; timing
        & \yescheck & \yescheck & \warntri & \yescheck & \nocross
        & \warntri & \nocross & \yescheck & \warntri & \yescheck
        & Equipment calibration, then attack \\
        \hline
        K\"oyl\"u et al.\newline ETS'24~\cite{koylu2024cgan}
        & VCC, EM & Conditional GAN; multilayer-perceptron classifier on the
          mute--pass boundary
        & $v,\ell,d$; EM $X$/$Y$, power, delay
        & \nocross & \yescheck & \nocross & \warntri & \yescheck
        & \yescheck & \nocross & \nocross & \nocross & \nocross
        & Glitch success rate; boundary \\
        \arrayrulecolor{ksBlack}\bottomrule
    \end{tabular}
    \par\vspace{2pt}
    \parbox{\textwidth}{\scriptsize\textcolor{ksCharcoal}{%
        \textbf{Legend:} \yescheck{}~present \quad \warntri{}~partial or
        limited \quad \nocross{}~absent \quad \nostated{}~not stated.
        Column polarity is normalized so that \yescheck{} always indicates the
        presence of a capability, which is why the first criterion is phrased
        as requiring \emph{no} prior fault map. The five prior works do not
        share an objective, so the marks position the contributions without
        ranking them.}}
    }
    \vspace{-12pt}
\end{table*}

A configuration may produce normal execution, a crash or reset, a missing or
invalid interaction, an incorrect but unusable result, or the desired target
fault.
Throughout the paper these five classes are tagged by color:
\OutcomeGreen{} (normal execution), \OutcomeYellow{} (crash, reset, or
synchronization timeout), \OutcomeMagenta{} (a non-target fault that disturbs
execution without meeting the target condition), \RED{}
(target fault), and \OutcomePink{} (missing trigger).
In practical campaigns, target faults often occupy comparatively small
and irregular regions of the explored space. These regions may be narrow or
disconnected, and their locations can depend on the device, firmware,
operating condition, and trigger alignment. Blind enumeration is consequently
expensive. Interpreting a response requires a program-specific outcome oracle. We write

\vspace{-7pt}
\begin{equation}
    o_t = \mathcal{O}(z_t,z_{\mathrm{ref}},m),
    \label{eq:oracle}
    \vspace{-9pt}
\end{equation}
\vspace{-7pt}

\noindent where $z_t$ is the response, $z_{\mathrm{ref}}$ is an
optional fault-free reference, and $m$ is the program-specific condition, such
as the AES signature, password-acceptance bit, or loop count
(Appendix~\ref{app:bench}). A non-target fault alters execution without meeting
this condition.
Structured outcomes can thus provide a graded search signal before the rare
target fault is first observed. Two conventional strategies ignore this accumulating signal.
Grid search discretizes each axis, so the number of attempts grows
multiplicatively with resolution. Random
search instead draws candidates from a fixed distribution over $\mathcal{X}$.
Neither method uses observed outcomes to reallocate future experiments. This
non-adaptive allocation is costly when the target region is sparse. A practical
search procedure should
instead use each hardware response to influence where it samples next, both to
find a useful region and to continue producing useful faults after discovery.

These two requirements give the two quantities we report. Under a budget of $B$
hardware attempts, the \emph{first-target iteration} is
$T_{\mathrm{target}}=\min\{t\leq B \mid o_t=o_{\mathrm{target}}\}$, the index
of the earliest attempt satisfying the target condition; it is right-censored
when no such attempt occurs within the budget.
It measures discovery cost in the resource that dominates a campaign. The
\emph{target-fault yield} is the rate at which the target outcome recurs from
that point onward,
$Y_B=(B-T_{\mathrm{target}}+1)^{-1}
\sum_{t=T_{\mathrm{target}}}^{B}\mathbbm{1}[o_t=o_{\mathrm{target}}]$, and is
undefined when $T_{\mathrm{target}}$ is right-censored. The two are distinct
objectives: a policy can reach a productive region quickly and then leave it, or
find it late and exploit it well, so neither quantity alone characterizes a
search procedure.

\subsection{Adaptive Parameter Search}
\label{sec:bg-optimization}\label{sec:bg-rl}

An online policy uses each observed outcome to focus subsequent
attempts. In a cold-start campaign it must learn entirely from attempts that
consume the budget, and a binary label is sparse before the first target fault;
the categorical outcome model of Section~\ref{sec:bg-campaigns} supplies
feedback beforehand.

Reinforcement learning (RL) provides one formulation for using this feedback
online~\cite{sutton2018rl}. A state summarizes the current
search context, an action selects the next configuration or search region, and
a reward quantifies the utility of the
observed outcome. Q-learning is a model-free algorithm for estimating the
long-run value of taking a given action in a given state purely from observed
transitions: after each step it nudges its current estimate toward the reward
just received plus a discounted estimate of the best value reachable from the
next state, so the estimate improves online without ever modeling the
environment's dynamics directly. In Q-learning, these decisions are represented by an
action-value function $Q(s,a)$ updated after each transition
$(s_t,a_t,r_t,s_{t+1})$~\cite{watkins1992qlearning},

\vspace{-7pt}
\begin{equation}
\scalebox{0.90}{$\displaystyle
    Q(s_t,a_t) \leftarrow Q(s_t,a_t)
    + \alpha\!\left[r_t + \gamma \max_{a'} Q(s_{t+1},a') - Q(s_t,a_t)\right]
$}
\label{eq:q-learning}
\vspace{-9pt}
\end{equation}
\vspace{-2pt}

\noindent where $\alpha$ is the learning rate and $\gamma$ discounts future
reward. Section~\ref{sec:design} specializes this formulation to FI and defines
the state, action, and reward used by our search policy.

\subsection{Prior Parameter-Search Work}\label{sec:bg-related}

Reducing the attempts needed to identify useful FI parameters has motivated
several alternatives to grid and random search. Boix Carpi et
al.~\cite{boixcarpi2013glitch} introduced Monte Carlo sampling, FastBoxing,
Adaptive Zoom and Bound, and genetic search for voltage glitching, exploiting
the observation that useful faults cluster near the transition between
configurations that leave execution unaffected and those that perturb it too
strongly. Later work extended evolutionary search with specialized crossover,
fitness, and memetic strategies~\cite{maldini2019optimizing}. These methods
avoid exhaustive enumeration, although their behavior can depend on the initial
population and on the regions found early.

Model-based methods instead learn from the configurations evaluated so far.
Bayesian optimization uses a surrogate and acquisition function to select
expensive black-box evaluations~\cite{jones1998ego,snoek2012bo,frazier2018bo,
bayesian_optimization}. Werner et al.~\cite{werner2021fast} formulated FI
equipment calibration as a hyperparameter-optimization problem and evaluated
SMAC, a random-forest-based configurator~\cite{hutter2011smac,breiman2001rf},
and successive halving~\cite{jamieson2016sha,li2018hyperband} over a
high-dimensional voltage-glitch space. Such surrogates are effective
when observed responses reveal a learnable region, but a cold-start campaign
must first obtain informative hardware outcomes from a space dominated by
normal execution or failed interactions. K\"oyl\"u et al.~\cite{koylu2024cgan} modeled the relation between outcome
classes and candidate configurations directly, using a conditional generative
model to propose samples near class boundaries. Together these results motivate
adaptive search and richer outcome representations, and none of them treats
first-target discovery, post-discovery reproduction, and fault-region
characterization as one objective.

We therefore study FI search from a cold start: the policy has no representative
fault map for the current device and program and must find its first target fault
within a limited hardware-attempt budget. After that discovery, the campaign
continues according to its objective, either reproducing the target fault
reliably or characterizing the broader vulnerable region. Each attempt reports
a structured outcome, such as normal execution, a non-target fault, a target
fault, or a failed interaction, and these distinctions provide search feedback
that a binary or scalar objective would discard. Section~\ref{sec:design}
develops two policies that use this feedback in complementary ways.

%% file: sections/threat_model.tex
\vspace{-7pt}

\section{Threat Model}\label{sec:threat}

We consider board-level voltage fault injection against embedded DUTs
without invasive access to the processor die. The same FI interface supports
two roles: an authorized security analyst evaluating device susceptibility and
a physical attacker seeking a security-relevant fault. Both select a glitch
configuration $\mathbf{x}=(v,d,\ell)$, as defined in
Equation~\ref{eq:glitch-configuration}, but may differ in target knowledge and
campaign objective.

\vspace{-12pt}

\paragraph{Security analyst.}
The analyst measures the DUT's fault susceptibility and produces evidence for
mitigation. In addition to the common capabilities defined below, the analyst
may use source code, firmware analysis, debug access, GPIO triggers, or power
traces to narrow the search. The campaign typically continues after the first
target fault to measure repeatability, map the vulnerable region, and identify
settings that countermeasures should cover. These knowledge sources improve
localization but are not required by the policies.
\vspace{-12pt}
\paragraph{Physical attacker.}
The attacker seeks a security-relevant effect after obtaining physical board
and supply access. The attacker controls the glitch parameters, repeatedly
invokes and resets the target operation, and observes an external response;
traces may also be used when probe access permits. The model does not require
source code, debug access, a prior fault map, or a pretrained target-specific
model. The attacker typically seeks the first target fault and then settings
that reproduce it reliably enough to support an attack.

\vspace{-2pt}
\subsection{Physical Capabilities}\label{sec:threat-capabilities}

We assume physical access to the DUT's PCB and an accessible processor supply
rail through which voltage glitches can be applied. The evaluator can control
glitch voltage, trigger-relative delay, and pulse duration $(v,d,\ell)$;
repeatedly invoke the target operation; observe its external response; and
reset the device between attempts. External measurements, such as power traces,
may be used for synchronization or to narrow the timing window, but are not
required by the optimizer.

We assume that a viable voltage-injection path has already been established;
identifying or engineering this path is outside the scope of this work.
Package decapsulation, laser injection, focused-ion-beam modification,
microprobing, and other die-level invasive techniques are excluded. Our focus
is not the physical generation of faults, but the efficient search within an
available FI parameter space.

\vspace{-3pt}
\subsection{Knowledge and Cold Start}
\label{sec:threat-knowledge}

Knowledge is treated independently from role, spanning white-box
access to source, firmware, and cooperative triggers through to black-box
access to the external interface alone. Across all settings, the search policy requires no previously characterized
parameter--outcome map or pretrained target-specific model. Each campaign
starts from explicit bounds on $(v,d,\ell)$ and learns from observations on the
current device and program. Source code, firmware analysis, traces, or protocol
timing may narrow the delay bounds; the measured campaigns use the localization
procedures in Section~\ref{sec:rlq-timing}. Thus, ``cold start'' means no
prior fault map, not an absence of all timing information.

Each hardware attempt is mapped by a program-specific oracle to a semantic
outcome class: normal execution, malformed or missing response, non-target
fault, target fault, or synchronization failure. The policy receives only the
tested configuration and this class; program-specific response parsing remains
inside the oracle.

\subsection{Objectives and Success Criteria}
\label{sec:threat-objectives}

The primary distinction between the two roles defined in
Section~\ref{sec:threat} -- the analyst and the attacker -- is the stopping
objective. An
attacker may prioritize the first security-relevant target fault, whereas an
analyst typically continues to measure reproducibility and characterize the
vulnerable region.
These objectives are the first-target iteration
$T_{\mathrm{target}}$ and the target-fault yield $Y_B$ of
Section~\ref{sec:bg-campaigns}, together with occupied parameter-space cells at
a controlled number of target faults.

We use \emph{fault-region discovery} to mean identifying a set of configurations
that repeatedly produce the target outcome, and not observing one
successful configuration. Accordingly, the evaluation reports discovery cost,
reproduction rate, and coverage as distinct quantities.

\paragraph{Scope.}
The model does not require source code, debug access, a prior fault map, a
pretrained target-specific model, or invasive die access. It requires only
repeatable invocation of the target operation, an established voltage-glitch
path, and an externally measurable outcome. We do not assume that every
supplied search space contains a successful fault region; rather, when one
exists, the problem studied here is how efficiently it can be discovered and
characterized.

%% file: sections/method.tex
\section{Proposed Search Methods and GlitchLab Framework}
\label{sec:method}\label{sec:design}
We present two policies that use the same semantic outcome classes but make
different structural assumptions. \RLQ{} exploits an ordered
voltage--pulse-length transition and treats delay as a gated search dimension.
\BOSOBAS{} (Structured-Outcome-Based Adaptive Search) instead learns
categorical outcome probabilities without an explicit physical threshold.
\GlitchLab{} executes both through the same closed-loop hardware interface,
allowing their candidate-selection policies to be compared under a common
experimental protocol.

\subsection{RL-Q: Physics- and Timing-Guided Search}\label{sec:rlq}
\vspace{-7pt}

\RLQ{} treats delay as a timing gate and voltage and pulse length as severity
controls, rather than searching three equivalent coordinates.

\subsubsection{Search Decomposition}\label{sec:rlq-decomp}


A useful fault requires timing overlap and suitable severity.
$\mathcal{W}$ is the set of delays for which the target operation is active;
thus, $d\in\mathcal{W}$ and $|\delta(v,\ell)|\leq\beta$.

The model is deliberately approximate: delay acts primarily as a gate, whereas
voltage and pulse length often provide ordered weak-to-severe feedback. Accordingly, \RLQ{} uses separate mechanisms for the three roles. It partitions
delay, uses tabular values with an upper-confidence bound (UCB) to allocate attempts
across amplitude--delay regions, and brackets pulse length within each selected
region. Each semantic outcome updates only the estimates for which it provides
evidence.

\subsubsection{Timing-Window Localization}\label{sec:rlq-timing}

Searching the full trigger-to-response interval is costly. \RLQ{} instead starts
from $\mathcal{D}_0=[d_{\min},d_{\max}]\supseteq\mathcal{W}$. Because host
scheduling and serial latency vary, the instrument anchors on a DUT hardware
edge at $T_{\mathrm{anchor}}$ and defines
$d=t_{\mathrm{glitch}}-T_{\mathrm{anchor}}$. This removes host latency while
retaining target execution-time variation (Section~\ref{sec:setup-sync}).

Figure~\ref{fig:rlq-delay}(a) previews how \RLQ{} subsequently refines the
resulting $\mathcal{D}_0$: it preserves full-range exploration, subdivides
target-producing segments, and stops at the timing-resolution floor.
In this figure, \RED{} denotes the target-fault outcome that
triggers subdivision. The two
routes below determine $\mathcal{D}_0$ before this online refinement begins.

\paragraph{Program and firmware knowledge.}\label{sec:rlq-delay-re} When the
image is available, a headless disassembler~\cite{ghidra} identifies the target
instruction and a GPIO edge is placed shortly before it. The attacked
instruction remains byte-identical, image size and layout remain fixed, and the
command path contains one rising edge. These conditions preserve the operation
and delay semantics, yielding a narrow $\mathcal{D}_0$. This route requires
image access and evaluates an instrumented binary.
This route is therefore available only to product-security
analysts.

\paragraph{Leakage-based localization.}\label{sec:rlq-delay-tvla} Without an
accurate program-level anchor, passive traces provide timing candidates. We
randomly interleave fixed and random executions and apply
non-specific Test Vector Leakage Assessment (TVLA)%
~\cite{tvla,leakage_assessment}. At sample $i$, the Welch statistic is
$t_i=(\bar{x}_{F,i}-\bar{x}_{R,i})/
\sqrt{s^2_{F,i}/n_F+s^2_{R,i}/n_R}$, using per-sample mean $\bar{x}$, variance
$s^2$, and size $n$ for the fixed ($F$) and random ($R$)
classes~\cite{welch}. Stable detections exceed a threshold with the same sign in
two data halves. They indicate class-dependent execution, not fault sensitivity,
so the exploration reserve in Section~\ref{sec:rlq-search}
may overrule them. A narrow support interval defines
$\mathcal{D}_0$; for broad periodic leakage, mean-trace
autocorrelation identifies repeated operations. Both routes consume measurement
time but no FI attempts.
This TVLA-based route applies to both analysts and attackers
when repeatable traces are available.

\subsubsection{RC-Informed Amplitude--Length Model}\label{sec:rlq-dose}\label{sec:design-dose}

Voltage and pulse length jointly set severity: a deep, short pulse and a
shallow, long one can produce similar excursions at the die, so useful faults
lie near the boundary between insufficient and excessive perturbation. A
single-pole rail driven from $V_{\mathrm{nom}}$ toward the commanded voltage $v$
reaches a critical level $V_c$ in
$D^{*}_{\mathrm{RC}}(v)=\tau_{\mathrm{rail}}\ln[(V_{\mathrm{nom}}-v)/(V_c-v)]$
for $v<V_c$, where $\tau_{\mathrm{rail}}$ is the effective injection-path time
constant. That relation motivates a monotone, convex strength--duration
boundary, but \RLQ{} assumes no knowledge of $\tau_{\mathrm{rail}}$, $V_c$, or
the injection network and approximates the observed trend online with

\vspace{-7pt}
\begin{equation}
    \widehat{D}(v) = c_2 v^2 + c_1 v + c_0.
    \label{eq:dose-curve}
    \vspace{-9pt}
\end{equation}
\vspace{-7pt}

The quadratic is an empirical approximation, fitted by weighted least squares
to a rolling window of faults per device, firmware, and injection path; no fit
transfers between benches. It summarizes the cross-amplitude trend but is not an
exact decision boundary. Each region therefore keeps an independent bracket and
local estimate, preventing noisy global curvature from shifting all regions.
The search coordinate is

\vspace{-7pt}
\begin{equation}
    \delta(v,\ell)=\ell-\widehat D(v).
    \label{eq:offset}
    \vspace{-9pt}
\end{equation}
\vspace{-7pt}

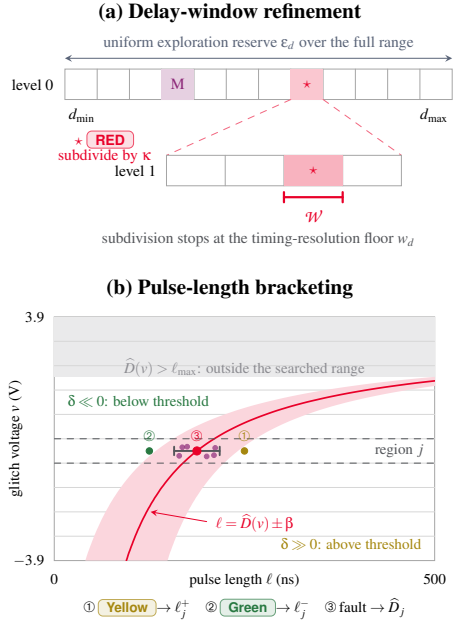
\begin{figure}[t]
    \vspace{-4pt}
    \centering
    \footnotesize
    \textbf{(a) Delay-window refinement}\\[2pt]
    \resizebox{0.7\columnwidth}{!}{%
    \begin{tikzpicture}[>=stealth, line cap=round]
        \def\W{6.1}
        \def\sw{0.50833}
        \draw[ksSlate,<->] (0,2.95) -- (\W,2.95);
        \node[ksSlate,font=\scriptsize,anchor=south] at ({0.5*\W},2.98)
            {uniform exploration reserve $\varepsilon_d$ over the full range};
        \draw[ksMedGray,fill=white] (0,2.30) rectangle (\W,2.78);
        \foreach \i in {1,...,11} \draw[ksMedGray] ({\i*\sw},2.30) -- ({\i*\sw},2.78);
        \fill[ksMagenta!22] ({3*\sw},2.30) rectangle ({4*\sw},2.78);
        \node[ksMagenta,font=\scriptsize] at ({3.5*\sw},2.54) {M};
        \fill[ksRed!22] ({7*\sw},2.30) rectangle ({8*\sw},2.78);
        \node[ksRed,font=\scriptsize] at ({7.5*\sw},2.54) {$\star$};
        \node[anchor=east,font=\scriptsize] at (-0.08,2.54) {level 0};
        \node[anchor=north west,font=\scriptsize] at (-0.04,2.26) {$d_{\min}$};
        \node[anchor=north east,font=\scriptsize] at ({\W+0.04},2.26) {$d_{\max}$};
        \draw[ksRed!60,dashed] ({7*\sw},2.30) -- (1.60,1.42);
        \draw[ksRed!60,dashed] ({8*\sw},2.30) -- (5.30,1.42);
        \node[ksRed,font=\scriptsize,anchor=east,align=center] at (1.45,1.55)
            {$\star$ \RED{}\\[-1pt]subdivide by $\kappa$};
        \draw[ksMedGray,fill=white] (1.60,0.94) rectangle (5.30,1.42);
        \foreach \i in {1,2,3} \draw[ksMedGray] ({1.60+\i*0.925},0.94) -- ({1.60+\i*0.925},1.42);
        \fill[ksRed!30] ({1.60+2*0.925},0.94) rectangle ({1.60+3*0.925},1.42);
        \node[ksRed,font=\scriptsize] at ({1.60+2.5*0.925},1.18) {$\star$};
        \node[anchor=east,font=\scriptsize] at (1.54,1.18) {level 1};
        \draw[ksRed,line width=1pt] ({1.60+2*0.925},0.76) -- ({1.60+3*0.925},0.76);
        \draw[ksRed,line width=1pt] ({1.60+2*0.925},0.68) -- ({1.60+2*0.925},0.84);
        \draw[ksRed,line width=1pt] ({1.60+3*0.925},0.68) -- ({1.60+3*0.925},0.84);
        \node[ksRed,font=\scriptsize,anchor=north] at ({1.60+2.5*0.925},0.66)
            {$\mathcal{W}$};
        \node[ksCharcoal,font=\scriptsize,anchor=north,align=center] at ({0.5*\W},0.30)
            {subdivision stops at the timing-resolution floor $w_d$};
    \end{tikzpicture}}
    \\[6pt]
    \textbf{(b) Pulse-length bracketing}\\[2pt]
    \resizebox{0.7\columnwidth}{!}{%
    \begin{tikzpicture}[>=stealth, line cap=round]
        \def\xs{0.01248}
        \def\ys{0.5128}
        \draw[ksMedGray] (0,0) rectangle (6.24,4.0);
        \fill[ksNeutral!18] (0,3.018) rectangle (6.24,4.0);
        \begin{scope}
            \clip (0,0) rectangle (6.24,4.0);
            \foreach \k in {1,...,9} \draw[ksMedGray!45] (0,{\k*0.4}) -- (6.24,{\k*0.4});
            \fill[ksRed!16]
                plot[domain=-3.9:2.05,samples=90,variable=\v,smooth]
                    ({(3490*ln((3.3-\v)/(3.107-\v))+54)*\xs},{(\v+3.9)*\ys})
                -- plot[domain=2.05:-3.9,samples=90,variable=\v,smooth]
                    ({(3490*ln((3.3-\v)/(3.107-\v))-54)*\xs},{(\v+3.9)*\ys})
                -- cycle;
            \draw[ksRed,thick]
                plot[domain=-3.9:2.05,samples=90,variable=\v,smooth]
                    ({3490*ln((3.3-\v)/(3.107-\v))*\xs},{(\v+3.9)*\ys});
            \draw[ksCharcoal,dashed] (0,1.600) -- (6.24,1.600);
            \draw[ksCharcoal,dashed] (0,2.000) -- (6.24,2.000);
            \foreach \p in {(2.05,1.72),(2.18,1.87),(2.50,1.71),(2.64,1.85),(2.62,1.73),(2.10,1.86)}
                \draw[ksMagenta,fill=ksMagenta,opacity=0.7] \p circle (1.1pt);
            \draw[ksCharcoal,line width=0.9pt] (1.966,1.800) -- (2.714,1.800);
            \draw[ksCharcoal,line width=0.9pt] (1.966,1.730) -- (1.966,1.870);
            \draw[ksCharcoal,line width=0.9pt] (2.714,1.730) -- (2.714,1.870);
            \draw[ksYellow,fill=ksYellow] (3.120,1.800) circle (1.5pt);
            \draw[ksGreen,fill=ksGreen] (1.560,1.800) circle (1.5pt);
            \draw[ksRed,fill=ksRed] (2.340,1.800) circle (1.9pt);
            \node[ksYellow,anchor=south,font=\scriptsize] at (3.120,1.860) {\ding{192}};
            \node[ksGreen,anchor=south,font=\scriptsize] at (1.560,1.860) {\ding{193}};
            \node[ksRed,anchor=south,font=\scriptsize] at (2.340,1.860) {\ding{194}};
            \node[ksCharcoal,anchor=east,font=\scriptsize] at (6.16,1.800) {region $j$};
        \end{scope}
        \node[anchor=north,font=\scriptsize] at (3.12,-0.12) {pulse length $\ell$ (ns)};
        \node[anchor=north,font=\scriptsize] at (3.12,-0.52)
            {\ding{192}\,\OutcomeYellow{}$\rightarrow \ell_j^{+}$\quad
             \ding{193}\,\OutcomeGreen{}$\rightarrow \ell_j^{-}$\quad
             \ding{194}\,fault $\rightarrow \widehat{D}_j$};
        \node[rotate=90,anchor=south,font=\scriptsize] at (-0.42,2.0)
            {glitch voltage $v$ (V)};
        \node[anchor=north,font=\scriptsize] at (0,-0.12) {0};
        \node[anchor=north,font=\scriptsize] at (6.24,-0.12) {500};
        \node[anchor=east,font=\scriptsize] at (-0.06,0) {$-3.9$};
        \node[anchor=east,font=\scriptsize] at (-0.06,4.0) {$3.9$};
        \node[ksGreen,font=\scriptsize,anchor=west] at (0.12,2.62)
            {$\delta \ll 0$: below threshold};
        \node[ksYellow,font=\scriptsize,anchor=east] at (6.12,0.28)
            {$\delta \gg 0$: above threshold};
        \node[ksNeutral,font=\scriptsize,anchor=north] at (3.12,3.42)
            {$\widehat{D}(v) > \ell_{\max}$: outside the searched range};
        \node[ksRed,font=\scriptsize,anchor=west] at (2.55,0.62)
            {$\ell = \widehat{D}(v) \pm \beta$};
        \draw[ksRed,->] (2.50,0.64) -- (1.58,0.80);
    \end{tikzpicture}}
    \caption{The two \RLQ{} search mechanisms. (a)~Delay-window
    refinement within $\mathcal{D}_0$: \RLQ{} retains uniform exploration over
    the initial window and subdivides target-producing segments; a segment
    producing only non-target faults (M) is not subdivided, and refinement stops
    at the timing floor $w_d$. (b)~Pulse-length localization within one
    amplitude region: the threshold locus $\widehat{D}(v)$ is monotone and
    convex, so at fixed voltage the outcome classes are ordered along the signed
    offset $\delta$ of Equation~\ref{eq:offset}. Probes \ding{192}
    and~\ding{193} set the bounds; bisecting between them places \ding{194}
    inside the transition, after which the region samples locally.}
    \label{fig:rlq-delay}\label{fig:rlq-dose}
    \vspace{-8pt}
\end{figure}

Negative values are below threshold, values near zero mark the fault
transition, and positive values are increasingly severe. Thus, \RLQ{} searches
one signed offset instead of unrelated voltage and pulse-length coordinates
(Figure~\ref{fig:rlq-dose}(b)).
\begin{figure*}[t]
    
    \centering
    \includegraphics[width=0.95\textwidth]{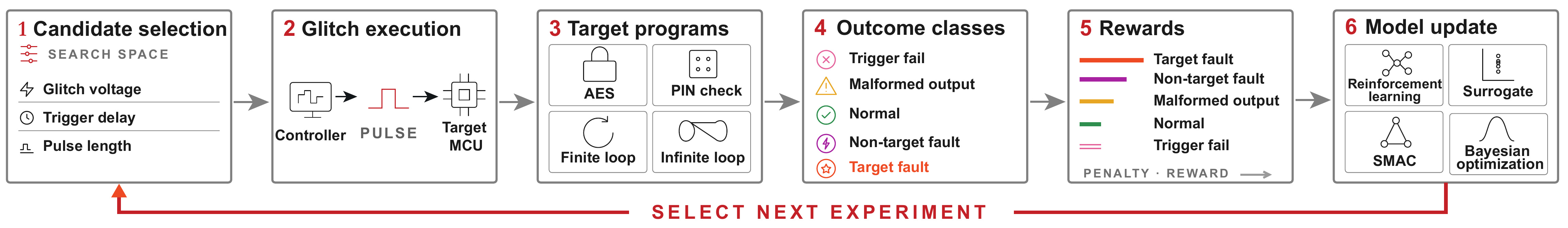}
    \vspace{-3pt}
    \caption{The \GlitchLab{} hardware-in-the-loop campaign. The active policy
    proposes $\mathbf{x}=(v,d,\ell)$, the campaign controller executes it on real
    hardware, the outcome oracle classifies the response, and the structured
    outcome returns to the policy. Only the policy block changes between the
    methods compared here.}\label{fig:glitchlab}
    \vspace{-10pt}
\end{figure*}

\subsubsection{Online Search and Allocation}\label{sec:rlq-search}

Each amplitude--delay region keeps a running average of the reward it has
produced so far, and every attempt is spent on the region that best balances a
high average against remaining uncertainty, so the search concentrates where
evidence suggests a fault is more likely without abandoning unexplored regions.


Action $a_t$ selects amplitude--delay cell $(j,k)$, and $Q_{jk}$ records its
utility. Each attempt starts from reset, so there is no next-state value to
bootstrap; prior outcomes still update the value table, threshold, and bracket.
Thus,

setting $\gamma=0$ in Equation~\ref{eq:q-learning} is a deliberate reduction to
a one-step structured bandit, yielding

\vspace{-7pt}
\begin{equation}
    N_{jk} \leftarrow N_{jk} + 1,
    \qquad
    Q_{jk} \leftarrow Q_{jk} + \frac{r_t - Q_{jk}}{N_{jk}},
    \label{eq:rlq-mean}
    \vspace{-4pt}
\end{equation}
\vspace{-7pt}

\noindent with Robbins--Monro rate $\alpha_t=1/N_{jk}$~\cite{robbins_monro}. The name
RL-Q refers to this tabular action-value update; we do not claim a long-horizon
Markov decision process. The table preserves the explicit region structure and
supports local decay when observations drift. \RLQ{} is therefore a structured
bandit and not a sequential decision process; the Q-value formulation is
retained because the region table, its optimistic initialization, and its
decay are naturally expressed in it.

\paragraph{Amplitude allocation.} The UCB table does not choose pulse length; it
picks which amplitude--delay region receives the next attempt, and that region's
bracket chooses pulse length. Amplitude is partitioned into $n_a$ equal-width
regions and $\mathcal{D}_0$ into $n_\phi$ delay bins, and,
following UCB~\cite{auer_ucb}, attempt $t$ selects

\vspace{-4pt}
\begin{equation}
    (j_t,k_t)
    =
    \arg\max_{j,k}
    \left\{
        Q_{jk}
        +
        c\,b_t
        \sqrt{\frac{\ln (t+1)}{N_{jk}+1}}
    \right\},
    \label{eq:rlq-ucb}
    \vspace{-3pt}
\end{equation}
\vspace{-4pt}

Here $N_{jk}$ counts visits, $c$ controls exploration, and $b_t$ increases it
during recovery; voltage and delay are uniform within the selected cell.
Optimistic $Q_{jk}\leftarrow q_0$, above any non-fault reward, makes an
unproductive first visit trigger coverage elsewhere. Regions with thresholds
above $\ell_{\max}$ are demoted by unperturbed responses but remain reachable.

\paragraph{Pulse-length localization.} At fixed amplitude, outcomes are roughly
ordered along pulse length, so the transition can be bracketed. Region $j$
maintains $[\ell_j^{-},\ell_j^{+}]$, the largest pulse seen on the weak side and
the smallest on the severe side. \OutcomeGreen{} raises the lower bound,
\OutcomeYellow{} lowers the upper bound, and \RED{}/\OutcomeMagenta{} update the
local transition estimate $\widehat D_j$:

\vspace{-7pt}
\begin{align}
    \OutcomeGreen{}:\;
    &\ell_j^{-} \leftarrow \max\!\left\{\ell_j^{-},\, \min(\ell,\, \ell_j^{+}-w_t)\right\},
    \label{eq:rlq-bracket}\\
    \OutcomeYellow{}:\;
    &\ell_j^{+} \leftarrow \min\!\left\{\ell_j^{+},\, \max(\ell,\, \ell_j^{-}+w_t)\right\},
    \nonumber\\
    \RED{}/\OutcomeMagenta{}:\;
    &h_j \leftarrow h_j+1,\quad
     \widehat{D}_j \leftarrow \widehat{D}_j + \frac{\ell - \widehat{D}_j}{h_j},
    \nonumber
    \vspace{-9pt}
\end{align}
\vspace{-7pt}

A fault recentres the bracket at $\widehat D_j\pm w_t/2$; the clamps enforce
a floor of $w_t$, preventing collapse to one length. Both fault classes mark the
transition and update $\widehat D_j$, although their rewards differ.

Before localization, \RLQ{} bisects the bracket with dither,
$\ell_t=\tfrac12(\ell_j^-+\ell_j^+)+u$ with
$u\sim\mathcal{U}(-W_j/8,\,W_j/8)$ and
$W_j=\ell_j^+-\ell_j^-$. Under the ordering assumption, each informative probe
roughly halves the interval. For $L=\ell_{\max}-\ell_{\min}$, localizing width
$2\beta$ takes $k^{\star}=\lceil\log_2(L/(2\beta))\rceil$
probes, versus $O(L/(2\beta))$ random samples; opening every region gives approximately
$n_a(1+k^\star)$ attempts. This finite-budget result covers only pulse length,
not the three-dimensional search.

After localization the bracket stops at a physical floor and samples
$\ell_t\sim\mathcal{N}(\widehat{D}_j,\sigma_j^{2})$ with
$\sigma_j=\sigma_\infty+(\sigma_0-\sigma_\infty)/(1+h_j/\nu)$. The initial width
is the transition half-width $\beta$, annealed toward $\sigma_\infty$ as
observations accumulate; the floor reflects measurement and timing uncertainty.
Either fault also enqueues an anisotropic Gaussian burst, wide on unlocalized
axes and tight on pulse length, with a longer burst after a target fault.

\paragraph{Delay refinement.} Pulse length gives directional feedback; delay
does not, since a timing miss does not say whether $\mathcal{W}$ lies earlier or
later and a non-target fault is no more reliable here. \RLQ{} therefore falls
back on the coarse coverage, segment scoring, and subdivision shown in
Figure~\ref{fig:rlq-delay}(a). By default the
$n_\phi$ bins enter Equation~\ref{eq:rlq-ucb} with widths no finer than the
measured execution-time spread; for a wide $\mathcal{D}_0$ the table factors
into marginal amplitude values $Q_a\in\mathbbm{R}^{n_a}$ and $m$ delay segments
scored by
\begin{equation}
    U_i
    =
    \underbrace{\frac{R_i}{n_i+1}}_{\text{target-fault rate}}
    +
    \underbrace{\eta\,\frac{M_i}{n_i+1}\cdot\frac{\tau}{\tau+n_i}}_{\text{decaying hint}}
    +
    c_d \sqrt{\frac{\ln (T+1)}{n_i}},
    \label{eq:rlq-delay-score}
    \vspace{-2pt}
\end{equation}

where $n_i$, $R_i$, and $M_i$ count attempts, target faults, and non-target
faults. The hint decays so repeated non-target faults cannot dominate; the last
term preserves exploration. A target-producing segment splits into $\kappa$
children while the resulting child width exceeds the timing
floor $w_d$.

Three safeguards prevent incorrect localization: an initial coarse sweep,
uniform full-range reserve $\varepsilon_d$, and partition rebuilding after a
fixed number of delay selections without a target fault. Counting only target
faults prevents non-target segments from trapping the search.

\subsubsection{Outcome Feedback and Adaptation}\label{sec:rlq-feedback}
\vspace{-7pt}

Each response updates only the model components it informs.
Each outcome updates only what it is evidence for: \OutcomeGreen{} and
\OutcomeYellow{} move the bracket bounds, \RED{} and \OutcomeMagenta{} move the
transition estimate, every valid class updates region utility, and
\OutcomePink{} affects failure handling alone. Region utility combines class
utility, proximity to the estimated transition, and execution latency:

\vspace{-7pt}
\begin{equation}
    r_t
    =
    R(o_t)
    -
    \beta_r
    \min\!\left(1, \frac{|\delta_t|}{w_t}\right)
    -
    \lambda \left( \frac{\tau_t}{\tau_{\min}} - 1 \right).
    \label{eq:rlq-reward}
    \vspace{-2pt}
\end{equation}
\vspace{-7pt}

$R(o_t)$ is the class utility (Table~\ref{tab:rlq-reward}) and the remaining
terms penalize threshold distance and latency; we set $\lambda=0$ because the
class utility already charges for outcomes needing recovery.
The \OutcomeYellow{} penalty marks a crash but excludes its full
reset time, which the evaluation reports separately.

A non-target fault is worth more before discovery, when it locates a
fault-producing transition, than after, when it consumes attempts without
meeting the objective, so \RLQ{} anneals $R_t(\OutcomeMagenta{})$ from
$r_M^{0}$ to $r_M^{\infty}$ over the first $T_M$ attempts. Physical drift makes
old observations less representative, so every $E$ attempts \RLQ{} applies
$Q\leftarrow\rho Q$ and $N\leftarrow\rho N$, giving effective memory
$E/(1-\rho)$, and after $S$ fault-free attempts it widens brackets. The policy
therefore tracks a region instead of fixing one configuration, which matters
when timing variation is comparable to the searched pulse-length interval.

\paragraph{Limitations.} The bracket assumes pulse-length outcomes are roughly
monotone at fixed amplitude; where that fails it can settle on the wrong pulse
length, and only widening and re-exploration recover. UCB regret guarantees also
assume stationary rewards~\cite{auer_ucb}, whereas annealing and decay make
\RLQ{} non-stationary. We therefore claim the finite-budget bisection efficiency
above, not asymptotic optimality of the whole algorithm. Algorithm~\ref{alg:rlq} summarizes the policy; delay comes from the joint grid
by default and from the partition of Equation~\ref{eq:rlq-delay-score} in
factored mode; Appendix~\ref{app:rlq-config} reports all values.


\begin{algorithm}[t]
\begin{minipage}{\columnwidth}

\caption{\RLQ{} physics- and timing-guided fault search.}
\label{alg:rlq}
\algbody
\fontsize{7}{8.7}\selectfont
\aline{0}{\kw{input:} bounds $\mathcal{X}$ with delay range $\mathcal{D}_0$; grid $n_a\!\times\!n_\phi$; reward table $R$}
\aline{0}{$Q\leftarrow q_0\mathbf{1}$; $N\leftarrow\mathbf{0}$; $w\leftarrow w_{\rm init}$; $s\leftarrow0$; initialize $[\ell_j^-,\ell_j^+]$, $h_j$}
\aline{0}{\kw{for} $j=1\dots n_a$: emit region-$j$ midpoint probe \cmt{open each bracket}}
\aline{0}{\kw{for} $t = 1, 2, \dots$ \kw{until} attempt budget exhausted}
\aline{1}{\kw{if} $t \bmod E = 0$ \kw{then} \textsc{Adapt}}
\aline{1}{\kw{if} local-burst queue non-empty \kw{then} $(v,d,\ell) \leftarrow$ pop}
\aline{1}{\kw{else}}
\aline{2}{$(j,k) \leftarrow \arg\max$ UCB \hfill (Eq.~\ref{eq:rlq-ucb})}
\aline{2}{$v \sim \mathcal{U}(\text{region } j)$;\;\; $d \sim \mathcal{U}(\text{bin } k)$}
\aline{2}{\kw{if} $\ell_j^+-\ell_j^->w$: bisect with
$\mathcal{U}(-W_j/8,\,W_j/8)$}
\aline{2}{\kw{else}: $\ell\sim\mathcal{N}(\widehat D_j,\sigma_j^2)$ \cmt{local sampling}}
\aline{1}{$o \leftarrow \textsc{Evaluate}(v,\, \mathrm{quant}(d),\, \mathrm{quant}(\ell))$ \cmt{one hardware attempt}}
\aline{1}{update bracket, $\widehat D_j,h_j$ (Eq.~\ref{eq:rlq-bracket}); refit $\widehat D(v)$ (Eq.~\ref{eq:dose-curve})}
\aline{1}{$r \leftarrow R(o) - \beta_r \min(1, |\delta|/w)$ \hfill (Eq.~\ref{eq:rlq-reward})}
\aline{1}{$N_{jk} \leftarrow N_{jk}+1$;\;\; $Q_{jk} \leftarrow Q_{jk} + (r-Q_{jk})/N_{jk}$}
\aline{1}{\kw{if} $o \in \{\RED{},\OutcomeMagenta{}\}$ \kw{then} enqueue $k_o$ local probes; $s \leftarrow 0$}
\aline{1}{\kw{else} $s \leftarrow s+1$}
\aline{0}{\kw{procedure} \textsc{Adapt}}
\aline{1}{\kw{if} $s > S$ \cmt{no recent fault: widen and re-explore}}
\aline{2}{$w \leftarrow \min(w_{\text{init}},\, w/\kappa_w)$;\; widen all brackets;\; $b\leftarrow 2$}
\aline{1}{\kw{else} $w \leftarrow \max(w_{\min},\, \kappa_w w)$;\; $b \leftarrow 1$}
\aline{1}{$Q \leftarrow \rho Q$;\;\; $N \leftarrow \rho N$ \cmt{decay}}
\algend
\vspace{-1pt}
\end{minipage}
\end{algorithm}

\subsection{SOBAS: Structured-Outcome Surrogate Search}
\label{sec:sobas}\label{sec:bo}\label{sec:bo-sobas}

\RLQ{} obtains efficiency from an amplitude--length ordering assumption. In
contrast, \BOSOBAS{} tests how far structured outcome labels can guide search
without that physical model. One random forest models all
outcome classes, whereas a second concentrates on the rare target event. The
acquisition combines target probability, failure suppression, and empirical
uncertainty.

\subsubsection{Structured Outcome Modeling}\label{sec:sobas-model}
\vspace{-7pt}


\BOGP{} and \BOSMAC{} reduce outcomes to one utility, so \OutcomeGreen{} and
\OutcomeYellow{} can receive similar scores despite giving opposite evidence.
\BOSOBAS{} instead models the classes separately.

\BOSOBAS{} instead learns the class-conditional distribution $P(o\mid\mathbf{x})$
with two random-forest classifiers~\cite{breiman2001rf} fitted in the unit cube
so no axis dominates a split. The multi-class forest predicts
$\hat{p}_k(\mathbf{x})=P(o=k\mid\mathbf{x})$ over the $K$ observed classes and
supplies the failure probabilities; the binary forest concentrates capacity on
the rare target event and predicts $\hat{p}_{\mathrm{t}}(\mathbf{x})$.
Disagreement across an evenly spaced tree subsample $\mathcal{B}$ gives an
uncalibrated empirical uncertainty $\hat{\sigma}_{\mathrm{t}}(\mathbf{x})=
\mathrm{sd}\{\hat{p}_{\mathrm{t}}^{(b)}(\mathbf{x})\}_{b\in\mathcal{B}}$.
Piecewise-constant forests suit the sharp, axis-dependent boundaries we
observe, where the evaluated GP represents them through a smooth kernel. Algorithm~\ref{alg:sobas} summarizes the policy. Unlike \RLQ{}, it learns the
geometry without an explicit estimator or partition. Its
complete configuration, including the scrambled Sobol warm-up, appears in
Appendix~\ref{app:sobas-config}~\cite{sobol1967sequence}.


\begin{algorithm}[t]
\begin{minipage}{\columnwidth}

\caption{\BOSOBAS{} structured-outcome search.}
\label{alg:sobas}
\algbody
\scriptsize
\aline{0}{\kw{input:} bounds $\mathcal{X}$ with delay range $\mathcal{D}_0$; warm-up size $n_w$; batch size $q$}
\aline{0}{$\mathcal H,\mathcal Q\leftarrow\emptyset$; evaluate $n_w$ Sobol warm-up points into $\mathcal H$}
\aline{0}{\kw{for} $t = n_w{+}1, n_w{+}2, \dots$ \kw{until} attempt budget exhausted}
\aline{1}{\kw{if} $\mathcal{Q} = \emptyset$ \cmt{refit and propose a new batch}}
\aline{2}{fit multi-class forest $\hat{p}_k$ and target forest $\hat{p}_{\mathrm{t}}$ on capped $\mathcal{H}$}
\aline{2}{\kw{if} no \RED{} observed in $\mathcal{H}$ \kw{then} $\alpha \leftarrow \alpha_{\mathrm{S}}^{0}$ \hfill (Eq.~\ref{eq:sobas-coldstart})}
\aline{2}{\kw{else} $\alpha \leftarrow \alpha_{\mathrm{S}}$ \hfill (Eq.~\ref{eq:sobas-acq})}
\aline{2}{$P\leftarrow$ scored uniform pool $\cup$ Gaussian neighbors of top points; rescore}
\aline{2}{$\mathcal{Q} \leftarrow$ top $q$ of $P$ after grid quantization and deduplication}
\aline{1}{$\mathbf{x} \leftarrow$ pop $\mathcal{Q}$;\; $o \leftarrow \textsc{Evaluate}(\mathbf{x})$ \cmt{one hardware attempt}}
\aline{1}{append $(\mathbf{x},o)$ to $\mathcal{H}$}
\algend
\vspace{1pt}
\end{minipage}
\end{algorithm}
\vspace{-7pt}

\subsubsection{Acquisition Function}\label{sec:sobas-acq}
\vspace{-7pt}

\BOSOBAS{} prefers candidates likely to produce the target, unlikely to crash or
miss the trigger, and uncertain enough to justify exploration, scoring their
product:

\vspace{-7pt}
\begin{equation}
    \alpha_{\mathrm{S}}(\mathbf{x})
    =
    \underbrace{\hat{p}_{\mathrm{t}}(\mathbf{x})}_{\text{exploit}}
    \cdot
    \underbrace{
        \bigl(1 - \hat{p}_{\mathrm{crash}}(\mathbf{x})\bigr)
        \bigl(1 - \hat{p}_{\mathrm{no\,trig}}(\mathbf{x})\bigr)
    }_{\text{constraints}}
    \cdot
    \underbrace{\bigl(\hat{\sigma}_{\mathrm{t}}(\mathbf{x}) + \epsilon\bigr)}_{\text{explore}},
    \label{eq:sobas-acq}
    \vspace{-4pt}
\end{equation}
\vspace{-7pt}

The multi-class factors suppress crashes and missing triggers; $\epsilon$ keeps
confident points revisitable. These are operational constraints, not a formal
safe-optimization guarantee.

\subsubsection{Cold-Start Exploration}\label{sec:sobas-cold}

Before the first target fault the binary model has no positive example and
cannot rank candidates. \BOSOBAS{} instead searches outcome boundaries by
normalized multi-class entropy, keeping the failure suppression:

\vspace{-7pt}
\begin{equation}
\begin{split}
    \alpha_{\mathrm{S}}^{0}(\mathbf{x})
    &=
    \bigl(1 - \hat{p}_{\mathrm{crash}}\bigr)
    \bigl(1 - \hat{p}_{\mathrm{no\,trig}}\bigr)
    \cdot
    \frac{H(\mathbf{x})}{\ln K},
    \\[2pt]
    H(\mathbf{x}) &= -\sum_{k=1}^{K} \hat{p}_k(\mathbf{x}) \ln \hat{p}_k(\mathbf{x}),
\end{split}
\label{eq:sobas-coldstart}
    \vspace{-9pt}
\end{equation}
\vspace{-7pt}

The policy switches to Equation~\ref{eq:sobas-acq} after its first target fault;
before then, Equation~\ref{eq:sobas-coldstart} governs discovery. Expected
improvement instead reduces to variance-seeking without a positive sample.

\subsubsection{Candidate Selection and Online Update}\label{sec:sobas-candidates}

Because a random-forest acquisition is piecewise constant, candidate selection
uses two stages. The policy first scores a uniform global pool and then scores
Gaussian perturbations around the highest-ranked candidates. It quantizes the
result to the hardware grid, removes duplicates, and evaluates the top $q$
configurations. Batching amortizes model fitting, while the capped history keeps
all recent observations and subsamples older attempts. This is a heuristic
acquisition; it does not provide an expected-improvement or regret guarantee.

\vspace{-7pt}
\subsection{\textsc{GlitchLab} Integration}\label{sec:glitchlab}\label{sec:design-framework}

\begin{table}[t]
    \centering\footnotesize
    \caption{Search bounds supplied to the policies. Voltage and pulse-length
    bounds are common to every target.}
    \label{tab:campaign-bounds}
    \begin{tabular}{@{}lrr@{}}
        \toprule
        Dimension or target & Cold span & Localized \\
        \midrule
        Voltage $v$ & \multicolumn{2}{c}{$-3.9$ to $3.9$\,V} \\
        Pulse length $\ell$ & \multicolumn{2}{c}{$0$ to $500$\,ns} \\
        \addlinespace
        \multicolumn{3}{@{}l}{\emph{Delay $d$ (ns)}} \\
        AES-128 & $1$--$49{,}500$ & $38{,}368$--$43{,}164$ \\
        Password check & $1$--$1{,}474$ & $40$--$80$ \\
        Finite loop & $0$--$5{,}000$ & $0$--$40$ \\
        \bottomrule
    \end{tabular}
    \vspace{-7pt}
\end{table}

For the controlled \RLQ{}--\BOSOBAS{} comparison, \GlitchLab{} fixes the
target, bounds, instrument sequence, oracle, recovery procedure, and attempt
budget. The resulting loop proposes a configuration, executes
one hardware attempt, classifies the response, updates the policy, and recovers
the target when necessary.

\vspace{-5pt}
\paragraph{Hardware campaign.}\label{sec:gl-loop}
Figure~\ref{fig:glitchlab} shows one budgeted attempt. Instrument completion
distinguishes a missing trigger (\OutcomePink{}) from a target-link timeout
(\OutcomeYellow{}).

\vspace{-7pt}
\paragraph{Interfaces.}\label{sec:gl-interfaces}
A policy proposes a configuration and receives its structured outcome. \RLQ{}
consumes the class, \BOSOBAS{} models class probabilities, and regression
policies receive one common scalar reduction. The instrument executes
$(v,d,\ell)$, obtains $z_{\mathrm{ref}}$ through a no-glitch path,
detects a completion timeout, and recovers from crashes. Oracle
$\mathcal O$ maps each program-specific response to the class and goal bit
defined by Equation~\ref{eq:oracle};
$\mathcal{O}_{\mathrm{AES}}$, $\mathcal{O}_{\mathrm{PIN}}$, and
$\mathcal{O}_{\mathrm{Loop}}$ encode their own success conditions, while
\OutcomeYellow{} and \OutcomePink{} retain fixed meanings. Retargeting changes
the instrument and oracle interfaces; policies need not interpret ciphertexts
or frames. We do not claim learned-policy transfer across programs.

\vspace{-5pt}
\paragraph{Common evaluation.}\label{sec:gl-baselines}

GP Bayesian optimization, SMAC, random search, and grid search use the same
campaign interface as reference policies~\cite{jones1998ego,snoek2012bo,
hutter2011smac,bergstra2012random}. The controller applies an identical
scalar outcome reduction to the two regression baselines
(Appendix~\ref{app:reference-config}). However, the current
reference campaigns
use the cold delay span (the full, unlocalized $\mathcal{D}_0=[d_{\min},d_{\max}]$
range listed under "Cold span" in Table~\ref{tab:campaign-bounds}), whereas
\RLQ{} and \BOSOBAS{} share localized delay
bounds. Consequently, only the \RLQ{}--\BOSOBAS{} comparison isolates candidate
selection; comparisons with reference policies evaluate complete campaign
configurations.

%% file: sections/experimental_setup.tex
\section{Implementation and Experimental Setup}\label{sec:impl}\label{sec:setup}

We describe the test bench, targets, search space, baselines, and protocol.
The setup keeps hardware execution, outcome classification, and attempt
accounting consistent across policies, enabling a controlled comparison.

\subsection{Test Bench and Timing Stability}
\label{sec:impl-instrument}\label{sec:impl-target}\label{sec:impl-host}%
\label{sec:setup-bench}\label{sec:setup-noise}\label{sec:setup-sync}

\begin{figure}[t]
    \centering
    \includegraphics[width=0.8\columnwidth]{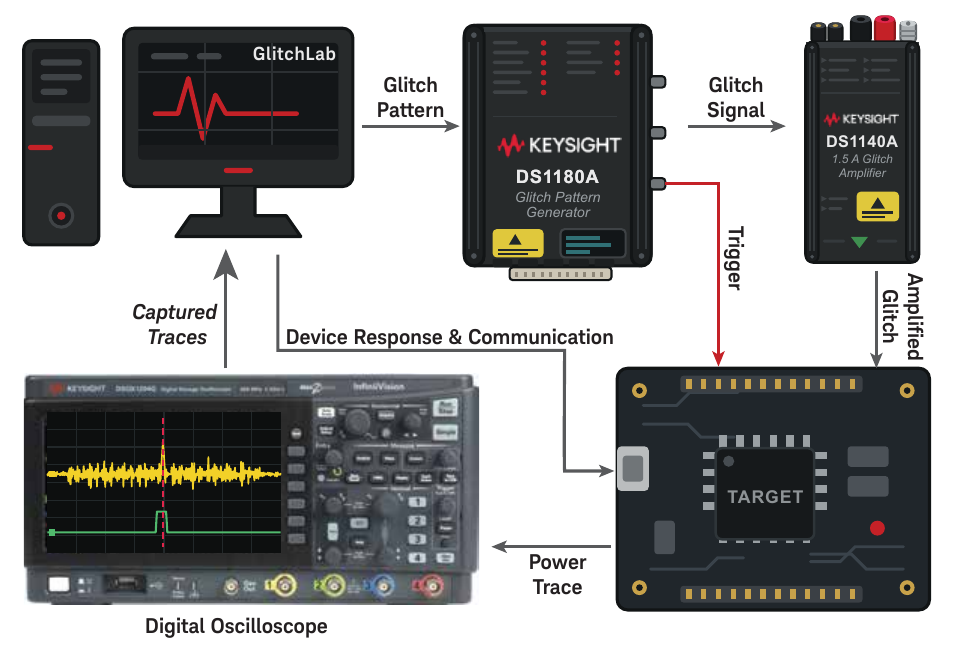}
    \vspace{-5pt}
    \caption{Representative evaluation bench used with each of
    three same-model Pinata boards. The generator and amplifier drive the
    target rail on a hardware trigger; UART returns responses and the
    oscilloscope captures localization traces.}
    \label{fig:exp-setup}
    \vspace{-8pt}
\end{figure}

Figure~\ref{fig:exp-setup} shows the representative evaluation bench used with
three separate units of the same Riscure Pinata model. A \ptag{ksCharcoal}{Keysight DS1180A} glitch generator and a \ptag{ksCharcoal}{Keysight DS1110A} glitch amplifier
drive the $3.3$\,V direct-to-die rail of each Pinata (STM32F407IG) running bare
metal at $168$\,MHz~\cite{stm32f407_datasheet,stm32f4_reference}. A Dell
Latitude 5520 (Core i7-1185G7, $32$\,GB, Ubuntu
24.04) runs the campaign controller and policy updates. The three units
test cross-unit repeatability under one injection path;
Section~\ref{sec:eval-threats} states what that does and does not establish.

To remove host-scheduling jitter, each attempt uploads an instrument-side state
machine that waits for the target's rising edge and, after $d$, drives the rail
to $V_f$ for $T_f$. An atomic bit-set generates the trigger; integer-nanosecond
timing quantizes $d$ and $\ell$ to $1$\,ns. Polled-UART request--response
protocols classify an unexpected response length as evidence of
perturbed execution, while the instrument
reports a missing trigger. Appendix~\ref{app:runtime} gives the
recovery and logging details. Across $30{,}000$ AES executions,
target timing within the $49{,}630$\,ns anchored window had a
standard deviation of $146$\,ns; this variation, and not instrument
resolution, is therefore the limiting uncertainty. Table~\ref{tab:campaign-bounds} gives the bounds supplied to every
campaign. Voltage and pulse-length bounds are common to all three targets. The
delay bounds are not, and the gap between the cold span and the localized
interval is why Section~\ref{sec:eval} treats the reference comparison as one
between campaign configurations and not between candidate-selection
policies.

\subsection{Search Space and Campaign Protocol}
\label{sec:setup-space}\label{sec:setup-protocol}\label{sec:setup-baselines}

The target conditions
are a four-byte corruption of standardized \AES{}~\cite{nist2023aes}, with a
signature motivated by byte-oriented differential fault
analysis~\cite{piret2003dfa},
acceptance of an incorrect password, and an early \FiniteLoop{} exit with a
parseable reply. Section~\ref{sec:security} gives their security
impact. All policies use the same voltage and pulse-length bounds,
oracle, reset procedure, timing quantization, and $5{,}000$-attempt budget.
For the controlled comparison, \RLQ{} and \BOSOBAS{} use the
same localized delay interval, whereas \BOGP{} and \BOSMAC{} use the
corresponding cold span.
Because \BOGP{} and
\BOSMAC{} consume scalar responses, the controller applies
the same scalar reduction to both and grades off-target faults by distance from
the four-byte signature (Appendix~\ref{app:reference-config}). \RLQ{} and \BOSOBAS{}
instead retain the semantic outcome class.

%% file: sections/evaluation.tex
\section{Evaluation}\label{sec:eval}

Fault-injection search is often summarized by one number: the time or number of
attempts required to obtain the first useful fault. We ask whether this measure
alone is sufficient, and find that it is not. The apparent leader therefore
depends on whether the campaign prioritizes discovery, reproduction, or
characterization. Table~\ref{tab:inversion} summarizes this result, and the
following subsections examine each objective in turn.

\begin{figure}[t]
    \centering
    \includegraphics[width=0.8\columnwidth]{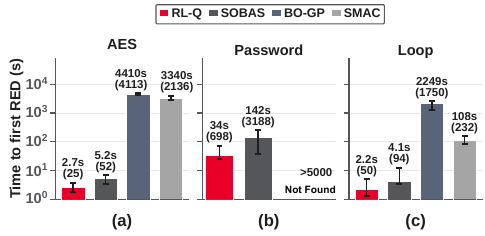}%
    \caption{Measured cost of the first target fault on (a) \AES{},
    (b) password and (c) the finite loop. Bars give median wall-clock time,
    with the median attempt index in parentheses; whiskers give the
    interquartile range across campaigns. Localization is excluded; $>5{,}000$ marks no target fault within the
    common attempt budget.}
    \label{fig:wallclock}
\end{figure}

\begin{table}[t]
    \centering\footnotesize
    \caption{Rank inversion across campaign objectives on the same bench,
    targets, and attempt budget.}
    \label{tab:inversion}
    \setlength{\tabcolsep}{3.5pt}
    \begin{tabular}{@{}llcr@{}}
        \toprule
        \textbf{Objective} & \textbf{Reported quantity} & \textbf{Leader}
        & \textbf{Margin} \\
        \midrule
        Discovery & Median attempts to first target
                     & \RLQ{} & $1.9$--$4.6\times$ \\
        Reproduction & Target-fault share after discovery
                     & \BOSOBAS{} & $7.3$--$21\times$ \\
        Coverage & New cells per target fault
                     & \RLQ{} & $4.8\times$ \\
        Coverage & Cells at equal yield
                     & \RLQ{} & $1.3\times$ \\
        \bottomrule
    \end{tabular}
    \vspace{-10pt}
\end{table}

\begin{figure}[t]
    \centering
    \includegraphics[width=0.8\columnwidth]{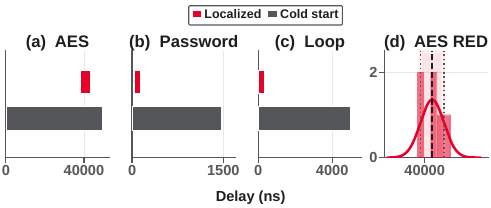}
    \caption{Delay localization. (a)--(c) Cold spans and
    assisted intervals for AES, password, and the finite loop. (d) Target-fault
    delays from an \protect\RLQ{} AES campaign initialized over the full span.}
    \label{fig:cold-win}
\end{figure}

Four questions organize the section: \ding{192}~how quickly each policy reaches
its first target fault; \ding{193}~how much inactive delay space localization
removes before adaptive search; \ding{194}~whether the measured
voltage--length geometry supports the ordering assumed by \RLQ{}; and
\ding{195}~how post-discovery yield trades against fault-region coverage. Each
subsection then relates the result to an evaluator characterizing a device and
an attacker seeking a repeatable effect.

\paragraph{Protocol.}
\RLQ{} and \BOSOBAS{} share the target, oracle, recovery path, search bounds,
and attempt budget; their comparison therefore isolates candidate selection.
\BOGP{} and \BOSMAC{} instead search the cold delay span in
Table~\ref{tab:campaign-bounds} and serve as reference configurations rather
than matched competitors. We execute ten independent campaigns
for every policy--target pair, including \BOGP{} and \BOSMAC{}, with fresh
policy state. We report medians and interquartile ranges descriptively and make
no inferential significance claims.

\subsection{First-Target Discovery}
\label{sec:eval-wall}

Figure~\ref{fig:wallclock} reports search-loop time and attempt index. Every
\RLQ{} and \BOSOBAS{} campaign succeeds on all three targets. On \AES{}, their
median costs are $25$ attempts/$2.7$\,s and $52$ attempts/$5.2$\,s. Relative to
SMAC, the faster successful reference configuration, these are $1{,}237\times$
and $642\times$ speedups. On the loop, medians of $50$ attempts/$2.2$\,s and
$94$ attempts/$4.1$\,s give $49\times$ and $26\times$ speedups. On password,
\RLQ{} and \BOSOBAS{}
succeed in median $698$ attempts/$34$\,s and $3{,}188$ attempts/$142$\,s,
whereas neither reference policy finds a target within $5{,}000$ attempts;
therefore, no finite speedup factor applies.

The proposed configurations are faster because they search a localized delay
interval and use structured outcomes through lightweight updates. \BOGP{} and
\BOSMAC{} search the full cold span and fit general-purpose surrogates across
many inactive configurations. Consequently, the proposed methods use
$2$--$85\times$ fewer hardware attempts, while model computation and differing
reset and recovery costs widen the post-localization wall-clock margin to
$26$--$1{,}237\times$.

The reference policies use cold-span bounds, so these factors
compare complete configurations; only the matched \RLQ{}--\BOSOBAS{}
comparison isolates candidate selection. \RLQ{} has
lower medians because its bracket can guide search immediately, whereas
\BOSOBAS{} needs a positive warm-up example. Their distributions overlap, so we
do not claim statistical separation.

\ResultBox{\ResultIconDiscovery}{\textbf{Discovery.} Both policies establish a
working fault in every campaign, so an evaluator can begin characterization and
an attacker can begin tuning within tens to hundreds of attempts, where the
reference configurations need thousands or fail outright.}

\subsection{Effect of Delay Localization}
\label{sec:eval-window}

Accurate delay localization aligns the glitch with the target operation and
avoids attempts outside the relevant execution interval. \GlitchLab{} supports
firmware-assisted GPIO anchoring, Grok 4.6-guided ranking of Test Vector Leakage
Assessment (TVLA) intervals, and \RLQ{} online refinement
(Section~\ref{sec:rlq-timing}).
Figure~\ref{fig:cold-win}(a)--(c) shows the resulting windows for \AES{},
password, and the finite loop; each is one to two orders of magnitude narrower
than its cold span. A focused hardware study in Appendix~\ref{app:llm-localization} evaluates a single fixed Grok 4.6 prompt. For two AES trace sets, Grok finds the target window in $18$ and $84$ attempts, compared with $129$ and $205$ attempts using magnitude-based TVLA ranking. This reduces the required validation attempts by $2.4$--$7.2\times$.


\RLQ{} can localize timing from campaign feedback. Equation~\ref{eq:rlq-delay-score}
uses each segment's target-fault rate and an exploration term, unlike uniform
random sampling. In Figure~\ref{fig:cold-win}(d), the first \AES{} target appears
after roughly six hundred attempts and later targets cluster nearby. Delay
locking retains this region; disabling it restores wider exploration.

\ResultBox{\ResultIconWindow}{\textbf{Localization.} The LLM-based workflow can
use GPIO or TVLA evidence to narrow the timing window, while \RLQ{} can recover
a target-producing interval directly from campaign feedback.}

\subsection{Fault-Region Geometry}
\label{sec:eval-vl}

\begin{figure}[t]
    \centering
    \includegraphics[width=\columnwidth]{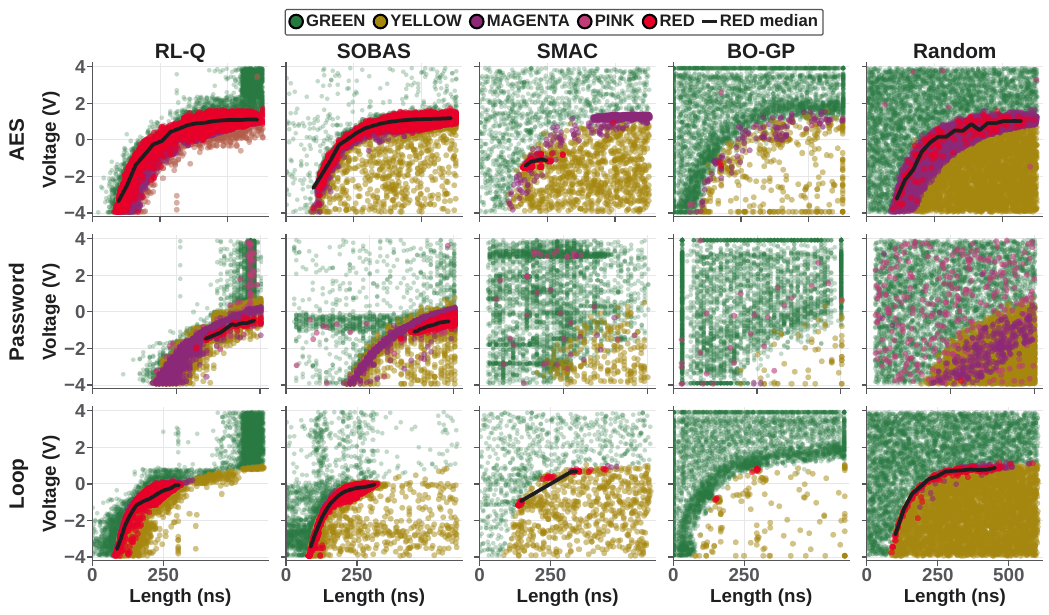}%
    \caption{Measured voltage--length outcome maps from
    hardware campaigns. Black curves are per-length target medians.}
    \label{fig:voltage-length}
\end{figure}

Useful faults lie between perturbations that are too weak and too severe.
Increasing voltage strength or pulse length crosses this band, as shown for all
three programs in Figure~\ref{fig:voltage-length}; delay must also reach the
target operation.

\begin{figure}[t]
    \centering
    \includegraphics[width=0.8\columnwidth]{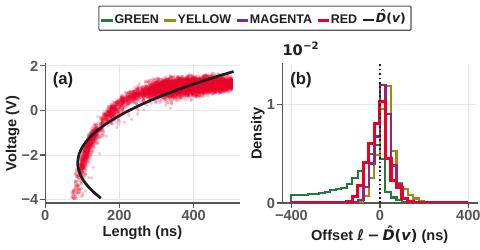}%
    \caption{Measured AES transition: (a) fitted $\widehat{D}(v)$ and
    (b) outcome density by offset. The fit uses $66{,}646$ targets
    ($R^2=0.74$).}
    \label{fig:threshold}
\end{figure}

Figure~\ref{fig:threshold} makes this transition explicit by fitting a quadratic
to the measured \AES{} target outcomes. The fit explains much of the
voltage--length variation, and expressing each attempt by its offset from the
curve straightens the boundary: target and off-target faults concentrate near
zero, while unperturbed and severe outcomes lie on opposite sides. The offset is
therefore useful as a search coordinate, but not as a deterministic classifier.
Delay, noise, and device state account for meaningful residual spread. The
amplitude--length fit is specific to the DUT rather than to \AES{} itself; only
the delay dimension is program-specific, so the password and loop results
provide only qualitative evidence of the same progression.

\RLQ{} encodes this ordering in its pulse-length bracket. In contrast,
\BOSOBAS{} makes no such assumption and instead learns a probability for each
outcome class from the random-forest classifiers of
Section~\ref{sec:sobas-model}. Focused seed-matched checks in
Appendix~\ref{app:llm-localization} show that removing the bracket delays one
AES discovery from $39$ to $189$ attempts, while replacing \BOSOBAS{}'s
structured model with a binary forest delays it from $54$ to $194$ attempts.
These single-seed checks illustrate mechanisms, not average
effects or confidence intervals.

\ResultBox{\ResultIconGeometry}{\textbf{Geometry.} The ordered transition gives
an evaluator a coordinate in which a vulnerable region can be described
compactly, and gives an attacker a direction along which a working setting can
be refined instead of rediscovered.}

\subsection{Repeatable Faults After Discovery}
\label{sec:eval-yield}

A single observation does not establish that a fault is reproducible. After
discovery, the relevant question is therefore how often the target outcome can
be obtained again. Here, the policies diverge. \BOSOBAS{} concentrates on
configurations that its model assigns a high target probability, whereas
\RLQ{} continues sampling along the transition to refine its boundary estimate.
As Table~\ref{tab:red-magenta} shows, \BOSOBAS{} reproduces the target outcome
roughly an order of magnitude more often on every program. This behavior is
valuable when an attacker must repeat a fault despite physical variability.
Conversely, \RLQ{} collects more off-target evidence near the boundary. Such
attempts contribute less to reproduction but more to characterization, as
Section~\ref{sec:eval-diversity} quantifies. The reference columns further illustrate this metric dependence. Yield alone
ranks \BOSMAC{} first on the loop but last on \AES{}, because the loop oracle is
less selective and rewards any policy that reaches a productive delay. Thus, a
ranking obtained from one metric reflects both the metric and the target
program; it should not be attributed to the search policy alone.

\begin{figure}[t]
    \centering
    \includegraphics[width=0.8\columnwidth]{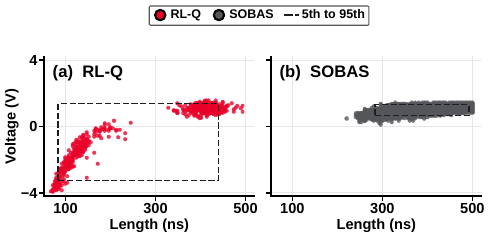}%
    \caption{Distinct AES target-fault cells for representative
    (a) \protect\RLQ{} and (b) \protect\BOSOBAS{} runs.}
    \label{fig:red-diversity}
\end{figure}

\input{plots/red_magenta_table}

\ResultBox{\ResultIconYield}{\textbf{Reproduction.} \BOSOBAS{} gives an attacker
a setting that repeats roughly an order of magnitude more often, while \RLQ{}
gives an evaluator the boundary evidence needed to place a guard band instead
of a single point.}

\subsection{Fault-Region Coverage}
\label{sec:eval-diversity}

Yield treats all target faults equally, although a fault at a previously unseen
setting provides more characterization evidence than another observation at a
known setting. We therefore use separate, longer \AES{} campaigns to measure
how much of the voltage--length plane each policy covers. Cells are coarser than
the timing grid so that near-duplicate settings do not count as distinct. Delay
is excluded because all runs share the same localized interval; this analysis
is specific to \AES{}.

Figure~\ref{fig:red-diversity} shows the difference directly: \RLQ{} follows the
transition across the voltage axis, whereas \BOSOBAS{} concentrates within one
pocket. Because \BOSOBAS{} produces many more target faults, its larger total
cell count is confounded by yield. Table~\ref{tab:red-diversity} therefore
compares coverage at equal yield. For the same number of target faults, \RLQ{}
covers more cells, spans several times more voltage, and converts a larger share
of its targets into previously unseen cells. This is the only comparison whose
interquartile ranges do not overlap, and the separation appears across three
coverage measures.

\input{plots/tab_red_diversity}

Taken together with Section~\ref{sec:eval-wall}, this result completes the rank
inversion in Table~\ref{tab:inversion}. The policy that reaches the first fault
earliest is not the one that reproduces it most often, and reproduction rate
does not indicate which policy characterizes the region most broadly. A
campaign must therefore report the measure that corresponds to its objective;
otherwise, a single ranking can obscure these distinct behaviors.

\ResultBox{\ResultIconCoverage}{\textbf{Coverage.} At equal yield \RLQ{} maps
more of the vulnerable region, which is what an evaluator needs to bound a
device's exposure, while \BOSOBAS{} concentrates where an attacker's next
attempt is likeliest to succeed.}

%% file: plots/red_magenta_table.tex
\begin{table}[t]
    \centering
    \scriptsize
    \setlength{\tabcolsep}{2.0pt}
    \caption{Share of attempts classified as \protect\RED{}
    (target fault) and \protect\OutcomeMagenta{} (off-target
    fault), pooled over measured campaigns. For \protect\AES{},
    \protect\RED{} is a four-byte difference on one ShiftRows-permuted
    MixColumns column (last-round DFA geometry); other four-byte
    patterns are counted with \protect\OutcomeMagenta{}.}
    \label{tab:red-magenta}
    \vspace{8pt}
    \resizebox{\columnwidth}{!}{%
    \begin{tabular}{@{}l rr rr rr rr rr@{}}
        \toprule
        & \multicolumn{2}{c}{RL-Q}
        & \multicolumn{2}{c}{SOBAS}
        & \multicolumn{2}{c}{SMAC}
        & \multicolumn{2}{c}{BO-GP}
        & \multicolumn{2}{c}{Random} \\
        \cmidrule(lr){2-3}\cmidrule(lr){4-5}
        \cmidrule(lr){6-7}\cmidrule(lr){8-9}
        \cmidrule(lr){10-11}
        & RED & Mag. & RED & Mag. & RED & Mag.
        & RED & Mag. & RED & Mag. \\
        \midrule
        AES & 6.4 & 17.2 & 60.2 & 10.6 & 0.01 & 58.8 & 0.01 & 1.1 & 0.15 & 4.4 \\
        Password & 0.16 & 3.6 & 3.4 & 12.3 & 0 & 0.14 & 0 & 0.04 & $<$0.01 & 0.61 \\
        Loop & 3.9 & 3.5 & 28.4 & 2.5 & 50.6 & 0.14 & 0.16 & 0.08 & 0.32 & 0.19 \\
        \bottomrule
    \end{tabular}}
    \vspace{-10pt}
\end{table}

%% file: plots/tab_red_diversity.tex
\begin{table}[t]
    \centering
    \scriptsize
    \setlength{\tabcolsep}{3.5pt}
    \caption{Diversity of AES \protect\RED{} settings in the
    voltage--length plane. }
    \label{tab:red-diversity}
    \begin{tabular}{@{}l rr@{}}
        \toprule
        Metric & RL-Q & SOBAS \\
        \midrule
        Unique $(v,\ell)$ cells & 456 [416, 492] & 952 [750, 989] \\
        Unique cells per RED & 0.72 [0.67, 0.75] & 0.15 [0.12, 0.17] \\
        Distinct voltage bins & 100 [94, 103] & 26 [22, 34] \\
        90\% RED voltage span (V) & 4.56 [4.27, 4.65] & 0.67 [0.60, 0.79] \\
        Unique cells after 400 REDs & 318 [305, 327] & 244 [164, 264] \\
        \bottomrule
    \end{tabular}
\end{table}

%% file: sections/security_impact.tex
\section{Security Impact of the Demonstrated Faults}
\label{sec:eval-security}\label{sec:security}

The preceding evaluation establishes how efficiently \GlitchLab{} finds and
characterizes target faults. We now connect those measured outcomes to their
security consequences. The three targets represent distinct attack primitives:
cryptographic corruption, authentication bypass, and control-flow violation.
They demonstrate security-relevant effects rather than complete end-to-end
exploits.

\textbf{Authentication bypass.} The password oracle records a target fault only
when the device accepts an incorrect password. The observed outcome therefore
constitutes a direct failure of the authentication decision. Both proposed
policies produce this outcome in every campaign, whereas \BOGP{} and \BOSMAC{}
do not produce it within the shared budget. Prior studies have similarly
bypassed PIN-verification logic and altered privilege decisions in embedded
systems~\cite{dutertre2021instruction,timmers2017linux}. Such a fault can expose
functionality protected by an access check.

\textbf{Control-flow violation.} The finite loop terminates before completing
the requested number of iterations, yet the program continues and returns a
parseable response with inconsistent counters. This behavior violates the
loop's intended control-flow guarantee and provides a primitive for skipping a
check, shortening a retry counter, or truncating a comparison. Prior work has
demonstrated instruction skips and program-counter corruption on 32-bit
microcontrollers, including redirection to attacker-selected
paths~\cite{moro2013emfi,timmers2016pc}. Our experiment establishes the
violation and the parameters that induce it, without inferring its underlying
microarchitectural mechanism.

\textbf{AES key recovery.} The \AES{} oracle accepts a ciphertext only when its
difference from the fault-free reference matches the geometry required for
last-round differential fault analysis: four differing bytes confined to one
ShiftRows-permuted MixColumns column. A small number of correct and faulty
ciphertext pairs is sufficient to recover the last-round key and invert the
AES-128 key schedule using established methods~\cite{dusart2003aes,piret2003dfa}.
The practical bottleneck is acquiring suitable pairs from hardware.
Table~\ref{tab:red-magenta} shows that \BOSOBAS{} produces them in a substantial
fraction of its attempts. We evaluate this acquisition step and rely on the
established literature for the subsequent deterministic key-recovery procedure. Showing DFA is out of scope of this paper. 

%% file: sections/discussion.tex
\section{Discussion and Limitations}\label{sec:discussion}\label{sec:eval-threats}


The central contribution is an online search that reaches target faults using
far fewer attempts than existing baselines by assigning distinct roles to
delay, voltage, and pulse length. The same structure can include parameters
from other modalities, such as EMFI position. Table~\ref{tab:inversion} also
shows that \RLQ{} favors discovery and coverage, whereas \BOSOBAS{} favors
reproduction. These objectives should be reported separately.

\textbf{Reporting FI campaigns.} A campaign should report three complementary
quantities.

\emph{Discovery cost} should be reported in both hardware attempts and
wall-clock time, which includes reset and recovery -- \textit{unlike a normal
or missing-trigger outcome, a crash or synchronization failure requires the DUT
to be power-cycled and re-armed before the next attempt can begin}. Figure~\ref{fig:wallclock}
reports both as medians and interquartile ranges across ten campaigns.

\emph{Reproduction rate} should be measured after the first target fault,
because one successful attempt does not establish repeatability. Finally,
\emph{region coverage at controlled yield} distinguishes broad exploration
from repeated sampling of one productive setting. Discovery cost without
reproduction rate cannot distinguish a repeatable vulnerability from an
isolated event; reproduction rate without coverage cannot distinguish a broad
vulnerable region from a narrow one.

\textbf{Accounting for localization.} A timing miss gives no
earlier-or-later information, although its class may indicate severity. Narrowing the range therefore increases the
information obtained from each subsequent attempt; in our experiments, it
removes one to two orders of magnitude of inactive delay space. \GlitchLab{}
supports firmware-assisted GPIO localization, LLM-coordinated TVLA, and
\RLQ{}'s online refinement. These routes incur different costs: firmware
instrumentation, trace acquisition and analysis, or additional hardware
attempts. Results obtained over different initial delay ranges therefore
compare complete campaign configurations, not optimizers alone. 

\textbf{Scope and limitations.} We evaluate three same-model Pinata boards,
three programs, and one voltage-injection path. This establishes cross-unit
repeatability, but not transfer across device models, architectures, production
firmware, or injection modalities. The fitted strength--duration constants are
bench-specific. Comparisons against cold-span references assess complete
campaigns; only the matched \RLQ{}--\BOSOBAS{} experiments isolate optimizer
behavior. Campaign-level results support descriptive
comparisons, while yield and coverage
remain within-study measures set by oracle selectivity and cell resolution.

These limits concern validation breadth, not dependence on prior data. Neither
policy requires a precomputed fault map or observations from another device;
both learn from the current target within user-defined bounds. They can
therefore be applied to new platforms and programs once the instrument
interface, search space, and target oracle are defined. Validating such transfer
remains future work.

\textbf{Future modalities.} Our evaluation focuses on voltage fault injection.
Future work will extend \GlitchLab{} to clock glitching, electromagnetic fault
injection, and laser fault injection. The same closed-loop formulation applies
when a modality exposes controllable parameters and informative outcome
feedback, although each extension requires modality-specific constraints and
experimental validation.

%% file: sections/conclusion.tex
\section{Conclusions}\label{sec:conclusion}

Voltage fault injection is an online search problem: a campaign must align a
physical perturbation with sensitive execution and distinguish the target
outcome from several other responses. \GlitchLab{} closes this loop while
preserving the semantic outcome classes and supports two policies with
complementary structural assumptions. Across three same-model units, both
policies find a target fault in every campaign. The reference configurations
fail on the password target and require substantially more attempts on the
other programs. After discovery, however, the ranking changes: \BOSOBAS{}
reproduces target faults more frequently, whereas \RLQ{} covers more of the
vulnerable region at equal yield. The resulting faults bypass authentication,
violate intended control flow, and provide the ciphertext differences required
for AES differential fault analysis. Although this evaluation focuses on voltage fault injection, the closed-loop formulation is not tied to voltage parameters. It can extend to other
fault-injection search spaces when modality-specific controls and informative
outcome feedback are available. Thus, even on identical hardware and under the same budget, the apparent policy
ordering depends on the campaign objective. Fault-injection studies should
preserve semantic outcomes and report discovery cost, reproduction rate, and
coverage separately. \GlitchLab{} facilitates efficient, objective-aware fault
search for security analysts mapping vulnerable regions and attackers
identifying repeatable fault settings on devices, without treating either goal
as a universal measure of search quality.

%% file: sections/appendix.tex
\raggedbottom

\section{Reproducibility and Experimental Details}
\label{app:repro}\label{app:params}
\vspace{-7pt}

This appendix follows a campaign from its shared experimental definition to
policy instantiation, physical execution, outcome classification, and logging.
This order separates parameters that must remain fixed across policies from
parameters that intentionally distinguish the search methods. The values
describe measured voltage-fault-injection campaigns, not simulations, and are
provided so that the reported distributions can be reproduced without access
to the implementation. Because every campaign starts from fresh stochastic
state, the reproducibility target is the campaign-level distribution and not
an identical sequence of proposed configurations.

\subsection{LLM-Assisted Delay Localization}
\label{app:llm-localization}
\vspace{-7pt}

\GlitchLab{} uses Grok 4.6 to turn Test Vector Leakage
Assessment (TVLA) output into an ordered set of candidate delay windows. One
temperature-zero prompt is frozen across all targets and trace sets. It receives
only the measured TVLA curve, its statistical summary, and a fixed window width;
it cannot inspect firmware, GPIO placement, previous FI campaigns, or manually
selected delay intervals. Grok 4.6 produced the rankings once before hardware
validation, and the campaign controller then replayed the frozen rankings. This
separation prevents target-fault observations from influencing window selection.

Across five hardware-localization cases, Grok 4.6 ranked the
target-producing interval first in both AES cases and
reduced cumulative validation from $129$ and $205$ attempts under magnitude-only
TVLA ranking to $18$ and $84$ attempts, respectively, a $2.4$--$7.2\times$
reduction. On password set~1, the Grok-ranked first window produced a target in
$164$ attempts, whereas magnitude ranking and full-span online search exhausted
$2{,}500$ and $5{,}000$ attempts. Password set~2 and the loop demonstrate why
\GlitchLab{} retains online refinement: \RLQ{} found the password interval missed
by both trace rankings, while the loop was found fastest by magnitude ranking.
The three routes therefore provide complementary coverage, while Grok-guided
ranking substantially reduces validation effort when broad or misleading TVLA
peaks obscure the target operation.

\begin{table*}[t]
    \centering
    \begin{minipage}[t]{0.59\textwidth}
    \color{black}\fontsize{7}{7.8}\selectfont
    \textbf{Frozen Grok 4.6 prompt (verbatim).}\par\smallskip
    \ttfamily
    You are given TVLA output for one Pinata program. Produce an ordered
    list of at most five candidate delay windows for a subsequent VCC glitch search.\par
    \smallskip
    Rules:\par
    - Read only the t-curve, tvla\_result.json summary, and the frozen window width.
      Do not use a hand-set delay window, firmware disassembly, GPIO placement, or
      any previous campaign result.\par
    - Every window has the same width W specified for this program.\par
    - Windows must not overlap.\par
    - If samples\_over\_threshold\_pct is above 80 AND the capture spans a multi-round
      operation (AES), the |t| map is saturated and does not localize. Autocorrelate
      mean\_a\_mv, take the dominant lag as the round period, and rank the last two
      repeating periods first (the second-to-last period, then the last), then
      earlier periods. The final period is often I/O; the previous one is the
      better candidate for a computational fault.\par
    - Do not treat a short cropped capture that is mostly over threshold as
      saturated periodicity. Password and loop stay on the narrow-support rule.\par
    - Narrow support: rank the earliest strong post-trigger local maxima
      (|t| $>$ 4.5), not the global |t| maximum if that maximum is late relative to
      the first instruction-scale lobe.\par
    - Exclude t $<$ 0. On AES also exclude the trigger falling-edge crosstalk (last
      \textasciitilde200 ns of the anchored window).\par
    - Amplitude stays -3.9 .. 3.9. Pulse length is not constrained by this ranking.\par
    \smallskip
    Return an ordered list of windows [d\_min, d\_max], method name, and a short
    rationale.
    \end{minipage}\hfill
    \begin{minipage}[t]{0.38\textwidth}
    \color{black}\scriptsize
    \textbf{Frozen configuration.}\par\smallskip
    \begin{tabular}{@{}ll@{}}
        Model & Grok 4.6 \\
        Temperature & $0.0$ \\
        Prompt SHA-256 & \texttt{c9c4d58bc3608c64deeb8107ab6d118b}\\
                       & \texttt{6f1dbe22c7cc79cb99dd7fcc4631ec71} \\
        Maximum windows & $5$, non-overlapping \\
        TVLA threshold & $|t|>4.5$ \\
        Widths & AES: $4{,}796$\,ns \\
               & password/loop: $40$\,ns \\
    \end{tabular}

    \end{minipage}
    \caption{\textcolor{black}{Frozen prompt and configuration for LLM-assisted
    delay localization. Candidate windows are edge-clipped at the capture
    boundary; therefore, a nominal $40$\,ns window may be shorter near $t=0$.}}
    \label{tab:focused-study}
\end{table*}

These results show that the main policy components make useful
faults easier to find. For AES, \RLQ{} finds the first target in $39$ attempts
with length bracketing, but needs $189$ attempts without it. \BOSOBAS{} finds
the first AES target in $54$ attempts with its multiclass model, compared with
$194$ attempts with a binary model; without entropy-guided initialization, it
finds no target within $1{,}500$ attempts. Semantic outcome classes make little
difference in the single AES run, but they are important for password bypass:
the complete \RLQ{} policy succeeds at attempt $115$, while the binary version
finds no target within $1{,}500$ attempts. Overall, these checks provide useful
evidence that the components improve discovery, although each comparison uses
only one matched campaign.

\begin{table}[H]
    \centering\footnotesize
    \caption{Campaign settings used for the reported hardware experiments.
    Policy state and fitted models are reinitialized for every campaign.}
    \label{tab:campaign-config}
    \begin{tabular}{@{}>{\raggedright\arraybackslash}p{0.39\columnwidth}
                       >{\raggedright\arraybackslash}p{0.53\columnwidth}@{}}
        \toprule
        Setting & Value \\
        \midrule
        Injection modality & voltage fault injection \\
        Configuration & $\mathbf{x}=(v,d,\ell)$: voltage, delay, pulse length \\
        Physical units & three same-model Pinata boards \\
        Controlled targets & AES-128, password check, finite loop \\
        Discovery budget & $5{,}000$ attempts per campaign \\
        Timing quantization & $d$ and $\ell$ rounded to $1$\,ns \\
        Voltage handling & continuous within the stated bounds \\
        Initial state & empty history; no fitted model or value table retained \\
        Recovery & fixed target reset after a fault or failed response \\
        Recorded result & configuration, semantic outcome, oracle metadata,
                          timing, reset flag, and policy state \\

        \bottomrule
    \end{tabular}
\end{table}

\subsection{Campaign Protocol and Shared Search Space}
\label{app:campaign}
\vspace{-7pt}

The unit of reproduction is one complete campaign. First, select a target and
its delay interval from Table~\ref{tab:campaign-bounds} in Section~\ref{sec:setup-space}; second, initialize one
policy with no retained history; third, execute the fixed controller and oracle
for the budget in Table~\ref{tab:campaign-config}; finally,
derive the relevant discovery, yield, or coverage metric from the resulting attempt log. The controlled
\RLQ{}--\BOSOBAS{} comparison keeps the target firmware, bounds, oracle,
recovery path, and budget identical. BO-GP and SMAC use the cold delay spans,
so comparisons against them evaluate complete campaign configurations and
not candidate selection alone.

Once the shared campaign is fixed, only candidate selection and policy state
change. The following tables instantiate Algorithms~\ref{alg:rlq}
and~\ref{alg:sobas} and define the scalar interface used by the measured reference
policies. Together with the shared bounds above, they specify the complete
online search procedure.

\begin{table}[H]
    \centering\footnotesize
    \caption{\RLQ{} outcome utilities and adaptation schedule. The non-target
    utility is annealed linearly over the first $T_M$ attempts.}
    \label{tab:rlq-reward}
    \begin{tabular}{@{}llr@{}}
        \toprule
        Symbol & Meaning & Value \\
        \midrule
        $R(\RED{})$ & target fault & $1.0$ \\
        $r_M^0\rightarrow r_M^\infty$ & non-target fault & $0.5\rightarrow0.2$ \\
        $T_M$ & non-target anneal horizon & $300$ \\
        $R(\OutcomeGreen{})$ & unperturbed execution & $-0.05$ \\
        $R(\OutcomeYellow{})$ & crash or reset & $-0.15$ \\
        $R(\OutcomePink{})$ & missing trigger & $0$ \\
        $\beta_r$ & normalized offset penalty & $0.1$ \\
        $\lambda$ & execution-latency penalty & $0$ \\
        $E$ & attempts between adaptations & $40$ \\
        $\rho$ & value and count decay & $0.98$ \\
        \bottomrule
    \end{tabular}
\end{table}

\subsection{Policy Instantiations}
\label{app:policy-config}
\vspace{-7pt}

\subsubsection{RL-Q Configuration}
\label{app:rlq-config}
\vspace{-7pt}

Tables~\ref{tab:rlq-search-params} and~\ref{tab:rlq-reward} instantiate
Algorithm~\ref{alg:rlq}. With $\gamma=0$ and $\alpha_t=1/N_{jk}$, each cell
stores its running mean utility instead of a discounted return. Pulse-length
settings scale with $L=\ell_{\max}-\ell_{\min}$, allowing the same bracket
schedule to be applied to a different pulse-length range. The joint
$n_a\times n_\phi$ table is the default; the factored parameters apply when
adaptive wide-range delay refinement from
Equation~\ref{eq:rlq-delay-score} is enabled.

\subsubsection{SOBAS Configuration}
\label{app:sobas-config}
\vspace{-7pt}

Table~\ref{tab:sobas-params} instantiates Algorithm~\ref{alg:sobas}. Inputs are
mapped to $[0,1]^3$ before fitting so voltage, delay, and pulse length enter the
forests on comparable scales. Each refit trains one multi-class forest for the
outcome landscape and one binary forest for the target event. Candidate batches
are quantized and deduplicated before execution; their members are then executed
sequentially so every completed hardware attempt remains observable to the
campaign log.

\begin{table}[H]
    \centering\footnotesize
    \caption{\RLQ{} allocation, bracketing, and delay-partition parameters.}
    \label{tab:rlq-search-params}\label{tab:method-params}
    \begin{tabular}{@{}>{\raggedright\arraybackslash}p{0.24\columnwidth}
                       >{\raggedright\arraybackslash}p{0.47\columnwidth}
                       >{\raggedleft\arraybackslash}p{0.18\columnwidth}@{}}
        \toprule
        Symbol & Meaning & Value \\
        \midrule
        \multicolumn{3}{@{}l}{\emph{Action-value allocation}} \\
        $n_a$ / $n_\phi$ & voltage regions / delay bins & $10$ / $5$ \\
        $q_0$ & optimistic initial value & $0.5$ \\
        $c$ & UCB exploration scale & $0.4$ \\
        $\gamma$ & inter-attempt discount & $0$ \\
        $\alpha_t$ & cell update rate & $1/N_{jk}$ \\
        $b_t$ & normal / recovery multiplier & $1$ / $2$ \\
        -- & sampling inside selected cell & uniform \\
        -- & initial bracket probes & one per voltage region \\
        \addlinespace
        \multicolumn{3}{@{}l}{\emph{Pulse-length bracket}} \\
        $w_{\mathrm{init}}$ / $w_{\min}$ & initial / final floor
                                                & $0.30L$ / $0.12L$ \\
        $\kappa_w$ & floor contraction & $0.94$ \\
        -- & pre-localization dither &
        \makecell[r]{$\mathcal U(-W_j/8,$\\[-1pt]
                     $W_j/8)$} \\
        $\sigma_0$ / $\sigma_\infty$ & local sampling width
                                                & $0.10L$ / $0.03L$ \\
        $\nu$ & sampling-width anneal constant & $4$ \\
        $k_{\RED{}}$ / $k_{\OutcomeMagenta{}}$ & queued local probes
                                                & $3$ / $1$ \\
        \addlinespace
        \multicolumn{3}{@{}l}{\emph{Factored delay refinement}} \\
        $m$ / $\kappa$ & coarse segments / split factor & $12$ / $4$ \\
        $w_d$ & minimum child-segment width & $50$\,ns \\
        $\eta$ / $\tau$ & non-target hint weight / decay & $0.35$ / $60$ \\
        $c_d$ & delay UCB scale & $0.6$ \\
        $\varepsilon_d$ & uniform full-range reserve & $0.10$ \\
        \bottomrule
    \end{tabular}
\end{table}

The warm-up is a scrambled Sobol design~\cite{sobol1967sequence}. At the history
cap, all recent observations are retained and the older portion is sampled
uniformly. The tree-wise standard deviation in
Equation~\ref{eq:sobas-acq} is used only to rank candidates; it is not treated
as a calibrated posterior interval.

\begin{table}[H]
    \centering\footnotesize
    \caption{\BOSOBAS{} model, acquisition, candidate-generation, and
    history-management parameters. Width $0.05$ is in normalized coordinates.}
    \label{tab:sobas-params}
    \begin{tabular}{@{}>{\raggedright\arraybackslash}p{0.59\columnwidth}
                       >{\raggedleft\arraybackslash}p{0.32\columnwidth}@{}}
        \toprule
        Setting & Value \\
        \midrule
        Models & multi-class RF $+$ target RF \\
        Trees per forest & $50$ \\
        Minimum samples per leaf & $2$ \\
        Trees used for $\hat{\sigma}_{\mathrm t}$ & $12$ \\
        Input normalization & $[0,1]^3$ \\
        Sobol warm-up $n_w$ & $64$ \\
        Proposal batch $q$ & $16$ \\
        Acquisition before first target & normalized class entropy \\
        Acquisition after first target & target probability product \\
        Exploration floor $\epsilon$ & $10^{-3}$ \\
        Uniform global candidates per refit & $5{,}000$ \\
        Highest-ranked global seeds & $30$ \\
        Gaussian neighbors per seed & $12$ \\
        Local Gaussian width per axis & $0.05$ \\
        Local candidates before deduplication & $360$ \\
        Maximum training history & $1{,}500$ \\
        \bottomrule
    \end{tabular}
\end{table}

\subsubsection{Reference-Policy Configuration}
\label{app:reference-config}\label{app:baselines}
\vspace{-7pt}

BO-GP and SMAC receive the same scalar reduction below, preserving an identical
controller-to-policy interface while changing only how candidates are selected:

\begin{equation}
    R_{\mathrm{s}}(o,c) =
    \begin{cases}
        1.0                                                   & o = \RED{},\\[2pt]
        0.6\,\exp\!\left(-\dfrac{(c-4)^2}{2\sigma_c^2}\right)  & o = \OutcomeMagenta{},\\[6pt]
        0.1                                                   & o = \OutcomeGreen{},\\[2pt]
        -0.2                                                  & o = \OutcomeYellow{},\\[2pt]
        -0.5                                                  & o = \OutcomePink{},
    \end{cases}
    \label{eq:bo-reward}
\end{equation}

Here $c$ is the corrupted-byte count and $\sigma_c=1.5$ bytes. Only the
non-target-fault row is graded: the bell centered on four corrupted bytes gives
partial credit to an AES fault close to the target signature. This scalar score
is used only by \BOGP{} and \BOSMAC{}; \RLQ{} and \BOSOBAS{} retain the semantic
outcome class.

\subsection{Physical Execution and Outcome Classification}
\label{app:bench}
\vspace{-7pt}

A proposed vector becomes one hardware observation through a fixed physical
interface. This subsection records that interface independently of the policy,
making explicit which behavior belongs to the bench, target program, oracle,
and recovery controller.

\subsubsection{Instrument Signals and Timing}
\label{app:signals}
\vspace{-7pt}

Table~\ref{tab:pins} defines the electrical signal directions, while
Table~\ref{tab:runtime-config} records the runtime settings held fixed across
campaigns. The passive amplifier exposes no software-readable transfer state;
the voltage action is therefore the commanded generator setpoint $v$, not an
inferred amplifier slew rate. The target anchor is asserted and cleared by one
atomic write to GPIOC bit~2, avoiding a read--modify--write sequence in the
timed region.

\begin{table}[H]
    \centering\footnotesize
    \caption{Instrument signal assignment. The amplifier receives the
    generator output but provides no read-back channel.}
    \label{tab:pins}
    \begin{tabular}{@{}lll@{}}
        \toprule
        Signal & Line & Direction \\
        \midrule
        Trigger in & GPIO 0 & target $\rightarrow$ instrument \\
        Trigger enable & GPIO 8 & instrument $\rightarrow$ target \\
        Target reset & GPIO 2 & instrument $\rightarrow$ target \\
        Glitch out & OUT1 & instrument $\rightarrow$ amplifier \\
        \bottomrule
    \end{tabular}
\end{table}

\begin{table}[H]
    \centering\footnotesize
    \caption{Fixed bench and runtime settings relevant to reproduction.}
    \label{tab:runtime-config}
    \begin{tabular}{@{}>{\raggedright\arraybackslash}p{0.42\columnwidth}
                       >{\raggedright\arraybackslash}p{0.50\columnwidth}@{}}
        \toprule
        Setting & Value \\
        \midrule
        Generator / amplifier & DS1180A / DS1110A \\
        Target rail & $3.3$\,V direct-to-die \\
        Target & STM32F407IG, bare metal \\
        Target clock & fixed $168$\,MHz \\
        Host & Core i7-1185G7, $32$\,GB, Ubuntu 24.04 \\
        Controller runtime & Python 3.11 \\
        Target transport & polled UART; no receive ISR \\
        USB-CDC & disabled by jumper \\
        FTDI latency timer & $1$\,ms (host default: $16$\,ms) \\
        Instrument timing grid & $1$\,ns \\
        SysTick & $1$\,kHz, unmasked \\
        Completion handling & instrument completion flag and host timeout \\
        Log durability & flushed after every completed attempt \\
        \bottomrule
    \end{tabular}
\end{table}

\subsubsection{Target Commands and Outcome Oracles}
\label{app:targets}
\vspace{-7pt}

Table~\ref{tab:frames} connects each target command to the evidence consumed by
its oracle. This distinction is important: a parseable but incorrect response
is a program-specific fault, whereas a timeout, malformed frame, or missing
trigger is classified by the common campaign controller.

\begin{table}[H]
    \centering\footnotesize
    \caption{Target interfaces and program-specific target conditions.}
    \label{tab:frames}
    \begin{tabular}{@{}>{\raggedright\arraybackslash}p{0.18\columnwidth}
                       >{\raggedright\arraybackslash}p{0.26\columnwidth}
                       >{\raggedright\arraybackslash}p{0.44\columnwidth}@{}}
        \toprule
        Target & Request & Parsed response and target condition \\
        \midrule
        AES-128 & opcode $+$ 16\,B plaintext & 16\,B ciphertext; target is the
        four-byte differential signature \\
        \addlinespace
        Password check & incorrect password & accept/reject result; target is
        acceptance \\
        \addlinespace
        Finite loop & opcode $+$ 16-bit trip count, MSB first & 6\,B:
        \texttt{0xA5}, residual count (2\,B), iteration count (2\,B),
        \texttt{0xA5}; target is a premature exit with inconsistent counters \\
        \bottomrule
    \end{tabular}
\end{table}

The AES oracle compares the ciphertext with an unglitched reference before
testing the four-byte signature. The loop firmware decrements the requested
16-bit count while incrementing a second \texttt{volatile} counter, then returns
both; only a framed response with counters inconsistent with the requested trip
count is a target. The password oracle marks only acceptance of the supplied
incorrect password as a target. These predicates prevent crashes and malformed
communication from being counted as security-relevant faults.

\subsubsection{Controller State, Recovery, and Logging}
\label{app:runtime}\label{app:host}
\vspace{-7pt}

The controller executes the same ordered sequence for every policy:
\begin{enumerate}[leftmargin=1.35em,itemsep=0pt,topsep=2pt,parsep=0pt]
    \item obtain the unglitched reference and request one configuration;
    \item arm the instrument-side state machine and invoke the target;
    \item combine instrument completion with the parsed response to classify
          the outcome;
    \item append the complete record and update the active policy; and
    \item reset the target after a faulty or failed response.
\end{enumerate}
Instrument completion distinguishes a missing trigger from a target-link
timeout. Fixing this sequence ensures that measured policy differences arise
from candidate selection and not from inconsistent recovery or classification.

\GlitchLab{} uses Python~3.11. \RLQ{} uses NumPy; \BOSOBAS{} uses NumPy,
scikit-learn, and SciPy; \BOGP{} uses PyTorch, BoTorch, and GPyTorch; and
\BOSMAC{} uses SMAC3 and ConfigSpace. These packages provide the numerical
implementations, while the behavior needed to reconstruct the experiments is
defined by the equations, update order, bounds, and parameter tables in this
appendix.

Each attempt log stores $(v,d,\ell)$, semantic class, corrupted-byte count, raw
input and output, timing, reset status, and active policy state; \RLQ{} state
includes the coefficients of Equation~\ref{eq:dose-curve} and the offset of
Equation~\ref{eq:offset}. Flushing after every attempt preserves data after a
later hardware failure and supports recomputation of the reported first-target,
yield, and coverage metrics without rerunning the bench.